\documentclass[
reprint,
nofootinbib,
amsmath,amssymb,
aps,
prx,
superscriptaddress
]{revtex4-2}

\usepackage[english]{babel}
\usepackage[letterpaper,top=2cm,bottom=2cm,left=3cm,right=3cm,marginparwidth=1.75cm]{geometry}
\usepackage{amsmath}
\usepackage{amssymb}
\usepackage{physics}
\usepackage{graphicx}
\usepackage{csquotes}
\usepackage{xcolor}
\usepackage[colorlinks=true, allcolors=blue]{hyperref}

\usepackage{thmtools, thm-restate}

\usepackage[normalem]{ulem} 

\newtheorem{theorem}{Theorem}%[section]
\newtheorem{proposition}{Proposition}%[section]
\newtheorem{lemma}{Lemma}%[section]
\newtheorem{corollary}{Corollary}
\newcommand{\doub}{\Sd_{(2)}}

\newcommand{\sdag}{s^\dagger}
\newcommand{\Sd}{S^\dagger}
\newcommand{\Qd}{Q^\dagger}
\newcommand{\charge}{c}
\newcommand{\s}{s}
\newcommand{\bi}{b^i}
\newcommand{\Rmax}{2R}
\newcommand{\pRmax}{R_{\rm max}}
\newcommand{\diam}{\text{diam}}

\makeatletter 
    
\renewcommand\onecolumngrid{% <<<<<<
\do@columngrid{one}{\@ne}%
\def\set@footnotewidth{\onecolumngrid}% <<<<<<<<<<<<<<<<
\def\footnoterule{\kern-6pt\hrule width 1.5in\kern6pt}%
}

\renewcommand\twocolumngrid{% <<<<<<
        \def\footnoterule{% restore rule
        \dimen@\skip\footins\divide\dimen@\thr@@
        \kern-\dimen@\hrule width.5in\kern\dimen@}
        \do@columngrid{mlt}{\tw@}
}%

\makeatother    
\begin{document}

\title{Locality forces equal energy spacing of quantum many-body scar towers}
\author{Nicholas O'Dea}
\email{nickodea@princeton.edu}
\affiliation{Princeton Center for Theoretical Science, Princeton University, Princeton, NJ 08544, USA}
\affiliation{Princeton Quantum Initiative, Princeton University, Princeton, NJ 08544, USA}
\author{Lei Gioia}
\email{lei.gioia.y@gmail.com}
\affiliation{Walter Burke Institute for Theoretical Physics, Caltech, Pasadena, CA, USA}
\affiliation{Department of Physics, Caltech, Pasadena, CA, USA}
\author{Sanjay Moudgalya}
\email{sanjay.moudgalya@gmail.com}
\affiliation{School of Natural Sciences, Technische Universit\"{a}t M\"{u}nchen (TUM), James-Franck-Str. 1, 85748 Garching, Germany}
\affiliation{Munich Center for Quantum Science and Technology (MCQST), Schellingstr. 4, 80799 M\"{u}nchen, Germany}
\author{Olexei I.  Motrunich}
\email{motrunch@caltech.edu}
\affiliation{Walter Burke Institute for Theoretical Physics, Caltech, Pasadena, CA, USA}
\affiliation{Department of Physics, Caltech, Pasadena, CA, USA}

\begin{abstract}
Quantum many-body scars are non-thermal eigenstates embedded in the spectra of otherwise non-integrable Hamiltonians. Paradigmatic examples often appear as quasiparticle towers of states, such as the maximally ferromagnetic spin-1/2 states, also known as Dicke states. A distinguishing feature of quantum many-body scars is that they admit multiple local “parent” Hamiltonians for which they are exact eigenstates. In this work, we show that the locality of such parent Hamiltonians strongly constrains the relative placement of these states within the energy spectrum. In particular, we prove that if the full set of Dicke states are exact eigenstates of an extensive local Hamiltonian, then their energies must necessarily be equally spaced. Our proof builds on recent results concerning parent Hamiltonians of the $W$ state, together with general algebraic structures underlying such quasiparticle towers. We further demonstrate that this equal-spacing property extends to local Hamiltonians defined on arbitrary bounded-degree graphs, including regular lattices in any spatial dimension and expander graphs. Hamiltonians with $k$-local interactions and a bounded number of interaction terms per site are also encompassed by our proof. On the same classes of graphs, we additionally establish equal spacing for towers constructed from multi-site quasiparticles on top of product states. For the towers considered here, an immediate corollary of the equal spacing property is that any state initialized entirely within the quantum many-body scar manifold exhibits completely frozen entanglement dynamics under \textit{any} local Hamiltonian for which those scars are exact eigenstates. Overall, our results reveal a stringent interplay between locality and the structure of quantum many-body scars.
\end{abstract}

\maketitle

\onecolumngrid

\tableofcontents

\section{Introduction}\label{sec:intro}

Generic isolated quantum many-body systems governed by a local Hamiltonian are expected to thermalize under time evolution, with simple initial states evolving to highly entangled thermal states.
A signature of this phenomenon is that highly excited eigenstates of a generic local non-integrable Hamiltonian are hypothesized to be themselves thermal states, by a celebrated conjecture known as the Eigenstate Thermalization Hypothesis~\cite{Deutsch1991quantum, Srednicki1994chaos, Rigol2008thermalization, DAlessio2016quantum}.
However, stimulated by the discovery of unusual dynamics in Rydberg atom simulator experiments~\cite{bernien2017probing}, a large body of recent work has revealed that non-integrable local Hamiltonians can host atypical, nonthermal eigenstates known as Quantum Many-Body Scars (QMBS) embedded within an otherwise thermal spectrum~\cite{Turner_2018quantum, Turner_2018weak, moudgalya2018nonintegrable, moudgalya2018entanglement, Shiraishi_2017, Khemani_2019, Lin_2019, Choi_2019, Surace_2020, Iadecola_2019, Papic_2022, Serbyn-Papic2021_review, Moudgalya-Regnault2021_review, Chandran-Moessner2022_review}.
The existence of QMBS can lead to certain simple states completely avoiding thermalization and exhibiting simple dynamics such as persistent oscillations, which have been observed in multiple experiments~\cite{bernien2017probing, bluvstein2021controlling, kao2021pumping, bluvstein2022processor, jepsen2022phantom, zhang2022superconducting, chen2022pulse, su2023bose,desaules2024computer,austinharris2025spinor}.
The subject of QMBS continues to fascinate despite its relative maturity and extensive progress, see Refs.~\cite{Papic_2022, Serbyn-Papic2021_review, Moudgalya-Regnault2021_review, Chandran-Moessner2022_review} for reviews.
Even the original PXP model for the Rydberg atom chain continues to surprise us, with recent experimental discoveries of unusual dynamical properties and scar-like features beyond the original $\mathbb{Z}_2$ scars~\cite{Mark2025observation}, as well as theoretical findings of new types of approximate and exact scars~\cite{giudici2023unraveling, zlatko2023briding, Kerschbaumer2025quantum, Kerschbaumer2025discrete, Ivanov2024volume, Ivanov2025exact}.
On the broader scale, new models and types of scars are still being discovered~\cite{Gotta2023AsymptoticScars, Desaules2022SchwingerScars, Kunimi2023AsymptoticScarsDMI,kunimi2025systematic,gotta2025open,hashimoto2026construction,Langlett_2022,agarwal2023bell,srivatsa2023mobility,dong2023disorder,udupa2023weak,mohapatra2024exact,chiba2024exact,mukherjee2025tensor,mestyan2025crosscap, tamura2022hopping,sanada2023multibody,sanada2024integrable,faugno2025density,desaules2023weakschwinger,ge2024nonmesonic,miao2025exactquantummanybodyscars, mohapatra2025unraveling}, and there are important open questions also about the available theoretical frameworks~\cite{Papic_2022, Serbyn-Papic2021_review, Moudgalya-Regnault2021_review, Chandran-Moessner2022_review, moudgalya2023exhaustive}.
On the theory side, numerous works have found a wide variety of lattice models exhibiting QMBS, revealing a rich landscape of possibilities that can occur in local Hamiltonians.
A particularly striking class of QMBS are ``condensates" of simple quasiparticles generated by some quasiparticle creation operator $\Qd$ on some appropriate vacuum $\ket{v}$: $\ket{\psi^p} \propto (\Qd)^p\ket{v}$~\cite{moudgalya2018nonintegrable, moudgalya2018entanglement, Iadecola_2019, Choi_2019, Schecter2019weak, Mark2020Eta, moudgalya2020eta}.
We will call such a set of states a \textit{tower} even without a particular Hamiltonian in mind.
Among these, the simplest is the spin-1/2 ferromagnetic tower that consists of the highest weight states of global $SU(2)$ symmetry, also known as \textit{Dicke states}~\cite{Dicke1954Coherence,marconi2025symmetricquantumstatesreview} in the quantum information literature.
This QMBS tower, which we will call the \textit{Dicke tower}, has been studied extensively in many works~\cite{Choi_2019, Mark2020Eta, odea2020from, pakrouski2020many, pakrouski2021group, moudgalya2023exhaustive,Kunimi2023AsymptoticScarsDMI}, and forms the starting point for understanding the structure of more complicated towers of QMBS.
All known models with QMBS towers have equal energy spacing between consecutive states in the tower.
Such equally spaced towers have a simple but important dynamical consequence: any superposition of states in the tower will exhibit periodic motion, which implies distinctive and eternal oscillations in observables. 
Furthermore, as discussed in Ref.~\cite{odea2025entanglement}, many scarred models in the literature have the striking property that superpositions of scar states do not have any entanglement dynamics whatsoever---all entanglement measures are frozen in time.
The mechanism for freezing is the existence of simple (i.e., a sum of single-site terms) Hamiltonians that have the scars as eigenstates with the same energies as the original many-body Hamiltonian~\cite{odea2025entanglement}.
As a consequence, if the scars are equally spaced in energy by, e.g., $S^z_{\rm tot}$, then any Hamiltonian that splits them equally in energy will also generate scar dynamics characterized by the same entanglement freezing.
Given these dynamical consequences of equal spacing, it is natural to ask whether this equal spacing property of scar towers was in a sense inevitable.
That is, if these states are eigenstates, must their energies \textit{necessarily} be equally spaced?
The answer is clearly no for non-local Hamiltonians, but might physical constraints like locality enforce the equal spacing of quasiparticle towers?
This question originated as an equal spacing conjecture for several scar towers in Ref.~\cite{moudgalya2023exhaustive}.
Broadly, QMBS were understood in that work to be precisely those states that admit \textit{multiple} local Hamiltonians that have them as eigenstates, which we will refer to here as \textit{parent Hamiltonians}.\footnote{Note that this differs from the conventional use of the name ``parent Hamiltonian,'' which usually refers to frustration-free Hamiltonians for which the given states are \textit{ground states} (see, e.g., Refs.~\cite{Perez-Garcia2007,RevModPhys.93.045003}). Parent Hamiltonians for QMBS are often referred to as simply QMBS Hamiltonians.
We will also use ``local Hamiltonian" and ``extensive local Hamiltonian" interchangeably to mean Hamiltonians that are an extensive sum of strictly local terms (when operators or Hamiltonians have strictly bounded supports, we will explicitly call them strictly local for clear differentiation).}
The exploration of parent Hamiltonians for QMBS had been initiated in numerous earlier works that systematically constructed many classes of Hamiltonians with a given set of QMBS as eigenstates~\cite{Shiraishi_2017, Mark2020Eta, moudgalya2020eta, odea2020from, pakrouski2020many, pakrouski2021group, ren2021quasisymmetry, ren2022deformed}, often leveraging numerical techniques for identifying extensive local parent Hamiltonians~\cite{qiranard2017,chertkovclark2018,Mark2020Eta, odea2020from}.
This culminated in Ref.~\cite{moudgalya2023exhaustive} proposing to consider---and proving in many cases---the set of generators for the entire algebra of parent Hamiltonians for a given set of QMBS. 
While this ``exhaustive" algebra approach provides the formal expression for the full algebra of parent operators --- local and non-local--- that have the QMBS as eigenstates, it however does not provide a concrete prescription for constructing the exhaustive set of \textit{local} Hamiltonians in this algebra.
Nevertheless, motivated by the structure of the generators of such algebras, Ref.~\cite{moudgalya2023exhaustive} put forth several conjectures on the structure of extensive local Hamiltonians in that algebra.
For certain towers of QMBS, this included an \textit{equal spacing conjecture} that \textit{all} extensive local Hamiltonians that contain those QMBS as eigenstates necessarily have them equally spaced in the spectrum.
The veracity (or not) of such an equal spacing conjecture would follow from a complete characterization of extensive local parent Hamiltonians for a set of QMBS.
Important progress was made in a recent work~\cite{gioia2025distinct} for the case of the $W$ state as a QMBS.
The $W$ state is one of the simplest long-range entangled states, consisting of a single particle on a lattice hopping with zero momentum~\cite{PhysRevA.62.062314}, and it has a rich theoretical~\cite{PhysRevA.98.062335,PhysRevA.62.062314,PhysRevA.75.052109,kieferova2024logdepth,gioia2024wstateuniqueground,wampler2025absorbing} and experimental~\cite{eibl2004experimental,haffner2005scalable,bernien2017probing,song2017ten} history.
It is also the first non-trivial state in the Dicke tower consisting of the Dicke states $\ket{W^p}$,\footnote{In a system with $N$ sites, the Dicke states $\ket{W^p}$ are often written as $\ket{D^{N}_p}$ in the quantum information literature} which can be viewed as multi-particle generalizations of the $W$ state [see also Eq.~(\ref{eq:Wdef})]. 
Reference~\cite{gioia2025distinct} explicitly wrote down the complete set of extensive local parent Hamiltonians for the $W$ state in one spatial dimension (1D), and proposed a finer classification of the different kinds of parent Hamiltonians.
Specifically, it showed that any extensive local parent Hamiltonian for the $W$ state can be written as a sum of strictly local Hermitian annihilators of the $W$ state (type I Hamiltonians), strictly local non-Hermitian annihilators of the $W$ state (type II Hamiltonians),\footnote{Despite being non-Hermitian, they are useful for decomposing certain Hermitian operators into sums of local annihilators~\cite{gioia2025distinct}.} and the total number operator which splits the $W$ state from the vacuum state. 
The total number operator cannot be represented as a linear combination of type I and type II operators and is hence called a type III Hamiltonian.
Reference~\cite{gioia2025distinct} also extended some of the results to the rest of the states $\ket{W^p}$.
However, the full characterization of extensive local Hamiltonians for each of the $p > 1$ cases remains an open problem.
In this work, we partially extend the local parent Hamiltonian characterization to the entire Dicke tower of QMBS. 
In particular, we prove the equal spacing conjecture of Ref.~\cite{moudgalya2023exhaustive} for such states, showing that \textit{all} extensive local parent Hamiltonians for the Dicke states in 1D have them as equally spaced eigenstates.
To achieve this, we do \textit{not} need the full characterization of extensive local Hamiltonians for the Dicke tower: the proof of the equal spacing conjecture is already enabled by results in Ref.~\cite{gioia2025distinct} about type III Hamiltonians for the $W$ state. 
We combine this with well-known results on the algebraic structures of such towers of QMBS, particularly the effects of vanishing iterated commutators involving ladder operators~\cite{batista2009spiral, naoyuki2020onsager, moudgalya2020eta,Mark2020Eta, tang2022multimagnon,hashimoto2026construction}. We also introduce a simple characterization of the action of annihilators of the $W$ state on the rest of the tower. As we will discuss in more detail, these ingredients together ensure the equal spacing of the Dicke states under local Hamiltonians.
Remarkably, our proof techniques also allow us to extend this equal spacing result to Dicke states on bounded-degree graphs and to more general towers of states composed of quasiparticles on top of a product state.
The remainder of this paper is organized as follows. In Sec.~\ref{sec:Dicke}, we introduce the Dicke states and prove the equal spacing conjecture in 1D.
In Sec.~\ref{sec:Dickegraphs}, we generalize our proof to prove equal spacing of Dicke states on general bounded-degree graphs. In Sec.~\ref{sec:general}, we extend the proof to families of generalized towers with multi-site quasiparticles.
We end with a discussion of open questions and future directions in Sec.~\ref{sec:discussion}.

\section{Illustrative example: 1D Dicke Tower}\label{sec:Dicke}
Our most general results are in Sec.~\ref{sec:general}.
In this section, we work through the simplest example that is covered by our framework: the 1D Dicke tower.
This allows us to first introduce concepts in a simpler setting before generalizing. The Dicke tower on a system with $N$ qubits takes the form 
\begin{equation}\label{eq:Wdef}
|W^p\rangle \propto \left(\sum_i \sdag_i\right)^p \ket{\overline{0}}
\end{equation}
with $\ket{\overline{0}} = \ket{0}^{\otimes N}$.
Here $\sdag$ creates a hardcore boson or ``particle": $\sdag \ket{0} = \ket{1}$ and $\sdag \ket{1} = 0$.
The Hermitian conjugate of $\sdag$, $\s$, annihilates particles: $\s\ket{0}=0$ and $\s\ket{1}=\ket{0}$.
The conventional $W$-state ($p=1$) is one of the states of this tower; explicitly, it takes the form $\ket{W} = \frac{1}{\sqrt{N}} \sum_i \ket{0..01_i0...0}$.
This tower can equivalently be viewed in terms of spin-$1/2$ degrees of freedom: $\ket{0} = \ket{\uparrow}$ and $\ket{1} = \ket{\downarrow}$, with $\sdag = \sigma^-$ and $\s = \sigma^+$. In this case, the Dicke tower corresponds to the ``maximally ferromagnetic" states $\left( \sum_i \sigma^-_i\right)^p \ket{\uparrow}^{\otimes N}$; these states are eigenvectors of $(\sum_{i} \sigma^x_i)^2+(\sum_{i} \sigma^y_i)^2+(\sum_{i} \sigma^z_i)^2$ with maximal eigenvalue. 

In this section, we prove that whenever the states in the 1D Dicke tower are eigenstates of a local Hamiltonian (i.e., an extensive sum of strictly local terms), they will have equally spaced energies. Surprisingly, we do \textit{not} need to assume that $H$ is Hermitian, only that it is local.

To define locality, it is useful to introduce the concept of ``range."
An operator string of the form $\sdag_{i_1} ...\, \sdag_{i_n} \s_{i_{n+1}} ...\, \s_{i_{n+m}}$ has range $R$, where $R$ is the smallest natural number such that $|i_j - i_k|< R$ for $j,k \in {1,2,...,n+m}$.
Such ``normal-ordered" operator strings form a complete basis for operators, and we will define the range of a Hamiltonian as that of the largest range string appearing in it. 

We note in passing that in the spin language, $R$ can be analogously defined for the familiar Pauli string operator basis, and the two definitions of $R$ will agree.
We direct the reader to Ref.~\cite{gioia2025distinct} for further discussion of the normal-ordered operator basis. 
More generally, it will be useful to define the ``diameter" of a subset of sites $X=\{i_1,...,i_t\}$ as $\diam(X) = 1+\max_{j,k \in \{1,...,t\} } |i_j - i_k|$.
We will also denote cardinality of $X$ by $|X|$ (in this example, $|X| = t$).

It is also important to distinguish the range $R$ from ``$k$" in the conventional usage of $k$-locality; the latter only refers to the cardinality of the support.
For example, both $\sdag_1 \sdag_2$  and $\sdag_1 \sdag_{10}$ are $2$-local, although the former has range $2$ and the latter has range $10$.
We will call a Hamiltonian $k$-local if all the operator strings it contains are at most $k$-local.

Our main result in this section is for a 1D chain geometry and reads:
\begin{theorem}[Dicke Tower]\label{th:dicke}~\\
    Given a $k$-local Hamiltonian $H$ of range $R$ and system size $N>4kR$, if $H \ket{W^p} = E_p\ket{W^p}$ for all $p \in 0,1,...,L$, then $E_p = \Omega+\omega p$ for some constants $\Omega,\omega$. 
\end{theorem}
The theorem follows from three propositions that are of interest in their own right.
\begin{proposition}[Decomposition of $H$]\label{prop:decomp}~\\
    If $\ket{W}$ is an eigenstate of a local Hamiltonian $H$ of range $R$, then $H$ can be decomposed as  
    \begin{equation}\label{eq:decomp}
        H = \Omega I + \omega \sum_i \sdag_i \s_i + \sum_{X:\diam(X)\leq \Rmax} h_X ~,
    \end{equation}
    where $h_X \ket{W} = h_X\ket{\overline{0}}=0$ and $h_X$ is supported entirely within $X$.
\end{proposition}

\begin{proposition}[Annihilation of finite fraction]\label{prop:densityann}~\\
    Given a maximum range $\pRmax$ and a collection of operators $h_X$ (each labeled by its support $X$), where $h_X \ket{W} = h_X\ket{\overline{0}}=0$, if
    \begin{equation}
        \left( \sum_{X:\diam(X)\leq \pRmax} h_X \right) \ket{W^p} = E'_p\ket{W^p} ~,
    \end{equation}
    for some energies $E'_p$, then $E'_p=0$ for $p\leq N/\pRmax$.
\end{proposition}

\begin{proposition}[Induction on annihilation]\label{prop:annall}~\\
    Given a $k$-local Hamiltonian $H$, if 
    $H \ket{W^q} = 0$ for $q \leq 2k$, then $H \ket{W^p}=0$ for all $p$.
\end{proposition}

Before the proofs, a few comments are in order. For Theorem~\ref{th:dicke}, if $H$ is Hermitian, then the constants $\Omega,\omega\in\mathbb{\mathbb{R}}$; for non-Hermitian $H$, the constants $\Omega,\omega$ can be in $\mathbb{C}$.
For Prop.~\ref{prop:decomp}, $h_X$ is not restricted to be Hermitian, even when $H$ is Hermitian (in fact, some non-Hermitian $h_X$ will appear in general).
This proposition is a consequence of results in Ref.~\cite{gioia2025distinct}.
We will use Prop.~\ref{prop:densityann} with the choice of $\pRmax=2R$.
For Prop.~\ref{prop:annall}, the main assumption is that $H \ket{W^q} = 0$ for $q \leq 2k$. In particular, we do \textit{not} need to additionally assume that $\ket{W^p}$ is an eigenstate of $H$ for $p>2k$; that follows as a consequence of Prop.~\ref{prop:annall}.
\subsection{Proof of equal spacing for the 1D Dicke Tower}
\subsubsection{\textit{Theorem~\ref{th:dicke}: Dicke Tower}}
Our assumptions are that the Hamiltonian $H$ is $k$-local and range $R$, the system size $N$ is at least $4kR$, and all the Dicke tower states are eigenstates of $H$: $H \ket{W^p} = E_p \ket{W^p}$.

In particular, since $H$ has $\ket{W}$ as an eigenstate, by Prop.~\ref{prop:decomp}, $H$ can be written as Eq.~(\ref{eq:decomp}). Applying this form to $H \ket{W^p} = E_p \ket{W^p}$ and rearranging shows that
\begin{equation}
    \left( \sum_{X:\diam(X)\leq \Rmax} h_X \right) \ket{W^p} = [E_p - (\Omega + \omega p)] \ket{W^p} \equiv E_p'\ket{W^p} ~.
\end{equation}
If we can show that in fact $E_p' = 0$, we will have proven the theorem.

The proof that $E_p'=0$ proceeds in two steps. 

Prop.~\ref{prop:densityann} immediately applies with $\pRmax = \Rmax$. Thus $E_p'=0$ for $p \leq N/(\Rmax)$.

Since $N>4kR$, then $E_p'=0$ for at least $p=0,1,...,2k$.
Thus Prop.~\ref{prop:annall} shows $E_p'=0$ for all $p$. \hfill $\blacksquare$

\subsubsection{\textit{Prop.~\ref{prop:decomp}: Decomposition of $H$}}
This proposition is a consequence of results in Sec.~III of Ref.~\cite{gioia2025distinct}, in particular Table I.
We will review some of these results in Sec.~\ref{sec:Dickegraphs} and Appendix~\ref{app:Hdecompgengraph} when discussing a generalization of this proposition from a 1D chain to an arbitrary bounded degree graph.

\subsubsection{\textit{Prop.~\ref{prop:densityann}: Annihilation of finite fraction}}

Any Hamiltonian that is local relative to an open chain will be local relative to the closed chain, so we will without loss of generality take a closed chain geometry for the system.
For ease of notation, we will define $H' = \sum_{X:\diam(X)\leq \pRmax} h_X $.
Our goal is to show that if $H' \ket{W^p} = E_p' \ket{W^p}$, then $E_p'=0$ for $p<N/\pRmax$.

First, observe that for a given $X$ a Schmidt decomposition of $\ket{W^p}$ takes the form
    \begin{equation}
        \ket{W^p} = \sum_{l=0}^{\min(|X|,p)}  f_{l}^{X,N} \ket{W^l}_X\otimes \ket{W^{p-l}}_{X^c} ~, \qquad
        f_{l}^{X,N} := \sqrt{\frac{\binom{|X|}{l}\binom{N-|X|}{p-l}}{\binom{N}{p}}} ~,
    \label{eq:flapprox}
    \end{equation}
where $X^c$ is the complement of $X$, and $\ket{W^l}_X$ is the $\ket{W^l}$ state defined on the region $X$ such that $\langle W^l\ket{W^l}_X$ = 1, and similarly for $\ket{W^{p-l}}_{X^c}$ on the region $X^c$.
    
Let us first prove the proposition for $\ket{W^2}$, which will serve as an illustrative example.
Inspired by the Schmidt decomposition in Eq.~(\ref{eq:flapprox}), applying $H'$ to $\ket{W^2}$ yields
    \begin{align}
        H^{\prime}\ket{W^2} = \sum_{X:\diam(X)\leq \pRmax}f_2^{X,N} h_X \ket{W^2}_{X} \otimes \ket{0}_{X^c} ~,
    \end{align}
    where we have used $h_X\ket{W}_X = h_X\ket{0}_X = 0$.
    If we now demand $\ket{W^2}$ to be an eigenstate of $H^{\prime}$ with energy $E'_2$, then
    \begin{align}
    \sum_{X:\diam(X)\leq \pRmax} f_2^{X,N} h_X \ket{W^2}_{X} \otimes \ket{0}_{X^c} = E'_2 \ket{W^2} \quad.
    \label{eq:W2energy}
    \end{align}
    For $E'_2 \neq 0$, there are basis states on the right hand side that are not present on the left hand side. In particular, consider a state with two $1$'s separated by at least $\pRmax-1$ zeroes: $\ket{10...010...0}$ with the padding of $0...0$ on either side being at least $\pRmax - 1$ long ($N\geq2\pRmax$ will guarantee the existence of such a state on a closed chain).
    The left-hand side of Eq.~(\ref{eq:W2energy}) only has terms where two $1$'s must be  outside of $X^c$ and hence within distance $\leq \pRmax-1$ of each other, which prohibits this state, but the right hand side for $E'_2 \neq 0$ contains this state. Thus the equality can only be true for $E'_2 = 0$.

    More explicitly, we can apply this $\bra{10...010...0}$ state to both sides of Eq.~\eqref{eq:W2energy}: 
    \begin{align}
        0=\frac{E'_2}{\sqrt{\binom{N}{2}}} ~.
    \end{align}
    We generalize this idea to all $\ket{W^p}$ with $p \leq \frac{N}{\pRmax}$.
    In this case
    \begin{align}
        H^{\prime} \ket{W^p}= \sum_{X: \text{diam}(X) \leq \pRmax} \sum_{l=2}^{p} f_{l}^{X,N} h_X \ket{W^l}_X \otimes \ket{W^{p-l}}_{X^c} = E_p'\ket{W^p}\quad.
        \label{eq:HprimeWp}
    \end{align}
    Now we apply the bra $\bra{10...0}_{R_{\text{max}}}^{\otimes p} \otimes \bra{0...0}_{N - R_{\text{max}} p}$ to the LHS and RHS to obtain
    \begin{equation}
    0 = \frac{E_p'}{\sqrt{\binom{N}{p}}} ~,
    \end{equation}
    since for any $X$ on the LHS the applied bra contains at most one particle in $X$, implying that $X^c$ contains at least $p-1$ particles and hence gives zero when overlapped with $\ket{W^{p-l}}_{X^c}$ with $l \geq 2$.
    So we have shown that $\ket{W^p}$ with 
    $p \leq \frac{N}{\pRmax}$
    has eigenvalue $E_p'=0$ under $H$.\hfill $\blacksquare$

\subsubsection{\textit{Prop.~\ref{prop:annall}: Induction on annihilation}}

Before proving Prop.~\ref{prop:annall}, it is useful to first note the following lemma for the raising operator $\Sd = \sum_i s_i^\dagger$, which shows that iterated commutators with $\Sd$ will kill local operators. We represent an $n$-fold iterated commutator of $O$ with $S^\dagger$ as $[[[[O,\Sd],\Sd],...],\Sd]_{n}$. 
The lemma is relatively well-known in the scars literature (with various forms of this lemma appearing in e.g.~\cite{tang2022multimagnon,moudgalya2020eta}), but we include a quick proof to give intuition.

\begin{lemma}[Nilpotent commutator]\label{lemma:commutator}~\\
    For any $k$-local operator $O$, $[[[[O,\Sd],\Sd],...],\Sd]_{2k+1}=0$.
\end{lemma}
\textit{Proof.} It is useful to use an operator basis that slightly differs from the normal-ordered operator basis.
First consider a Hilbert space consisting of a single qubit. Introduce $s^z = \frac{1}{2}[\sdag,\s]$. Then $[\sdag,s] \propto s^z$, $[\sdag, s^z]\propto \sdag$, and $[\sdag,\sdag]=[\sdag,I]=0$. Since $I,\sdag,\s,s^z$ form a basis for single-site operators, the operation $[\sdag, \cdot]$ is a nilpotent operation that will return zero after at most three applications. For example, up to proportionality constants, the action of the commutator with $\sdag$ on $s$ sends $\s \to \s^z \to \sdag \to 0$.

Define the restriction of $\Sd$ to a region $X$ as $\Sd_X = \sum_{i \in X}\sdag_i$. Then, similarly to the above, the operation $[\Sd_X,\cdot]$ will return zero after $2 |X|+1$ applications to any operator since $(\Sd_X)^{|X|+1}=0$. 

Finally, a $k$-local operator $O$ is supported on some (possibly non-contiguous) region $X$ with $|X|=k$. 
Since $\Sd$ is a sum of single site operators, $[\Sd,A] = [\Sd_{X},A]$ and the support of the commutator is still within $X$. Thus a $(2k+1)$-fold iterated commutator with $\Sd$ will annihilate $A$. \hfill $\blacksquare$

The proof of Prop.~\ref{prop:annall} then proceeds via strong induction on the statement $H\ket{W^n}=0$. 

In particular, since we assume $H\ket{0}=H\ket{W}=...=H\ket{W^{2k}}=0$, we only need to prove the statement ``$H\ket{W^{n+1}}=0$ given that $H\ket{W^m}=0$ for $m=0,\dots,n$" for $n+1 > 2k$. Furthermore, since the set of states $\ket{W^p}$ ends at $p=L$, we only need to consider $n+1 \leq L$. 

First, up to a normalization constant $c_k$, 
\begin{equation}\label{eq:wpow}
    H \ket{W^{n+1}} = c_k H (\Sd)^{n+1} \ket{\overline{0}} ~.
\end{equation}
Now consider the $(n+1)$-fold iterated commutator $[[[[H,\Sd],\Sd],...],\Sd]_{n+1}$.
Expand out this commutator.
The commutator differs from $H (\Sd)^{n+1}$ only by terms where $H$ has fewer than $n+1$ factors of $\Sd$ to its right.
Since $H (\Sd)^m \ket{\overline{0}} \propto H\ket{W^m} = 0$ for $m<n+1$, this means 
\begin{equation}\label{eq:iterated}
    H(\Sd)^{n+1} \ket{\overline{0}} = [[[[H,\Sd],\Sd],...],\Sd]_{n+1} \ket{\overline{0}} ~.
\end{equation}
However, since $n+1>2k$, the iterated commutator vanishes, as $H$ can be decomposed into a linear combination of $k$-local operators.
Hence,
\begin{equation}
    H(\Sd)^{n+1} \ket{\overline{0}} = 0 ~.
\end{equation}

The proof thus proceeds by strong induction. \hfill $\blacksquare$

\section{Generalization to graphs}\label{sec:Dickegraphs}
In the subsequent parts of the paper, we will keep the graph structure of the system relatively general; we will only need the constraint that the number of nearest neighbors of any site is at most $\Delta$ for some $\Delta$ independent of system size.
To keep the notation light, we will still index sites using a single index $i$, and we will distinguish the location of such a site through $\mathbf{r}_i$.
We will define ``range" similarly to that in the 1D case: 
An operator string of the form $\sdag_{i_1} ...\, \sdag_{i_n} \s_{i_{n+1}} ...\, \s_{i_{n+m}}$ has range $R$, where $R$ is the smallest natural number such that $|\mathbf{r}_{i_j} - \mathbf{r}_{i_k}|< R$ for $j,k \in {1,2,...,n+m}$. We use $|\mathbf{r}_l - \mathbf{r}_m|$ to denote the graph distance between sites $l$ and $m$, where graph distance is the minimum number of edges traversed in order to travel from site $l$ to $m$. We define the diameter of a subset of sites $X$ analogously.

It will also be useful to define ``balls" on graphs: for each site $i$, define the ball $B_i(r)$ as the set of sites that are within graph distance $r$ of site $i$ (this includes those sites that are exactly graph distance $r$ away); we will also refer to $i$ as the center of the ball $B_i(r)$.
For example, $B_i(1)$ corresponds to the site $i$ and its nearest neighbors. We will use freely that on a graph with maximum vertex degree $\Delta$, the number of spins $|B_i(r)|$ that can fit in a ball of radius $r$ is bounded as
\begin{equation}
|B_i(r)| \leq1+\Delta+\Delta (\Delta-1) + \Delta(\Delta - 1)^2 + ... + \Delta (\Delta-1)^{r-1} = 1+\Delta \frac{ (\Delta-1)^r-1}{\Delta-2}\leq\Delta^r+1.
\end{equation}
In the later parts of the paper, we will also use the following well-known geometrical fact about balls on graphs. We include the proof for completeness in Appendix~\ref{app:geomproofs}.

\begin{lemma}[Sphere packing on graphs]\label{lemma:sphere}
On a graph with $N$ vertices and maximum vertex degree $\Delta$, it is possible to fit at least $N/(\Delta^{2r}+1)$ disjoint balls of radius $r$.
\end{lemma}

The lower bound in Lemma~\ref{lemma:sphere} on the number of disjoint balls can be quite loose; on Euclidean space of $D$ dimensions, it is possible to make the bound $N/(C_D r^D+1)$ for some geometric factor $C_D$.
For example, on a 1D line, the bound can be made $N/(2r+1)$.
However, the fact that a lower bound exists that takes the form $N/f(r)$ for a finite-valued function $f(r)$ is enough for our purposes; we will use this when proving Prop.~\ref{prop:densityanngengraph} and Prop.~\ref{prop:Q2class}.

Section~\ref{sec:general} will generalize Theorem~\ref{th:dicke} to more complicated towers on bounded degree graphs. As warmup, we will first extend Theorem~\ref{th:dicke} for the Dicke tower on bounded degree graphs in this section.

\begin{proposition}[Decomposition of $H$]\label{prop:decompgengraph}~\\
    On a graph with $N$ vertices and bounded degree $\Delta$, if $\ket{W}$ is an eigenstate of a local Hamiltonian $H$ of range $R$, then $H$ can be decomposed as  
    \begin{equation}\label{eq:gendecomp}
        H = \Omega I + \omega \sum_i \sdag_i \s_i + \sum_{X:\diam(X)\leq \Rmax} h_X ~,
    \end{equation}
    where $h_X \ket{W} = h_X\ket{\overline{0}}=0$ and $h_X$ is supported entirely within $X$.
\end{proposition}
\textit{Comment.} We implicitly assume that the graph contains three sites that are pairwise separated by a distance of more than $R$ and two sites that are pairwise separated by a distance of more than $2R$. This is guaranteed so long as $N > \Delta^{4R}+1$.

This proposition is a direct extension to arbitrary graphs of results in Sec.~III of Ref.~\cite{gioia2025distinct}, in particular Table I there.
Indeed, as noted in Ref.~\cite{gioia2025distinct}, most of the arguments leading to Table I there use only a weaker notion of locality akin to $k$-locality, which needs to be supplemented only by the notion/existence of sites that are sufficiently apart from each other relative to the interaction range $R$.
For completeness, in Appendix~\ref{app:Hdecompgengraph} we go over these arguments highlighting points particular to general graphs.

\begin{proposition}[Annihilation of finite fraction]\label{prop:densityanngengraph}~\\
Given a maximum range $\pRmax$ and a collection of operators $h_X$ (each labeled by its support $X$), where $h_X \ket{W} = h_X\ket{\overline{0}}=0$, if
    \begin{equation}
        \left( \sum_{X:\diam(X)\leq \pRmax} h_X \right) \ket{W^p} = E'_p\ket{W^p} ~,
    \end{equation}
    for some energies $E'_p$, then $E'_p=0$ for $p\leq N/(\Delta^{2 \pRmax}+1) $.
\end{proposition}

For the argument, we only need to replace the used $p$-particle product state from the one-dimensional case, $\ket{\Psi_{\text{prod;1D}}} = \ket{10 \dots 0}_{\pRmax}^{\otimes p} \otimes \ket{0 \dots 0}_{N - p\pRmax}$, with a construction appropriate for a more general graph. 

By examining an analog of Eq.~(\ref{eq:HprimeWp}),
all $p$-particle basis states in $\left( \sum_{X:\diam(X)\leq \pRmax} h_X \right) \ket{W^p}$ will have at least two particles within a region of diameter $\pRmax$. It suffices to exhibit a $p$-particle $\ket{\Psi_{\text{prod;gen}}}$ for which no two of the $p$ particles fit within a region of diameter $\pRmax$. This will be possible whenever $p$ disjoint balls of radius $\pRmax/2$ can fit on the graph. By Lemma~\ref{lemma:sphere}, this holds for $N>p(\Delta^{ \pRmax}+1)$.
Since such a $p$-particle configuration is present on the R.H.S.\ of Eq.~(\ref{eq:HprimeWp}), we conclude that $E'_p = 0$. \hfill $\blacksquare$

\begin{theorem}[Dicke tower on graphs]\label{th:dickegraphs}
On a graph of maximum degree $\Delta$, and given a $k$-local Hamiltonian $H$ of range $R$ and system size $N>2k(\Delta^{4R}+1)$, if $H \ket{W^p} = E_p\ket{W^p}$ for all $p \in 0,1,...,N$, then $E_p = \Omega+\omega p$ for some constants $\Omega,\omega$. 
\end{theorem}

The proof is exactly analogous to that of Theorem~\ref{th:dicke}, using Props.~\ref{prop:decompgengraph},~\ref{prop:densityanngengraph} in place of Props.~\ref{prop:decomp},~\ref{prop:densityann}. We can still use Prop.~\ref{prop:annall}, as it only used $k$-locality and not the underlying graph structure. \hfill $\blacksquare$

Our main assumption on the graph is that the degree is bounded, which is easily satisfied. For example, any geometrically local graph in higher dimensions has bounded degree (e.g. a cubic lattice). Much more exotic families of graphs also enjoy a bounded degree; for example, the kinds of expander graphs used in the construction of powerful quantum error correcting codes have bounded degree~\cite{sipser1996expander,
hastings2021fiber,panteleev2021degenerate,panteleev2021quantum,panteleev2022asymptotically,breuckmann2021balanced,leverrier2022quantum,dinur2023good}.
We highlight that our result is useful even when there is not an (obvious) underlying graph structure. For example, consider a $k$-local Hamiltonian such that every site is part of at most $T_{\text{max}}$ terms in the Hamiltonian. Such a ``low-density" Hamiltonian actually induces a graph structure: define a graph where two sites are connected by an edge so long as there is a term in the Hamiltonian that involves both sites. The degree of any site on this graph is thus at most $(k-1)T_{\text{max}}$. Furthermore, with respect to this bounded-degree graph, the Hamiltonian terms have range $R=2$. Thus low-density Hamiltonians are covered by Theorem~\ref{th:dickegraphs}.

\section{Generalized towers}\label{sec:general}

In Sections~\ref{sec:Dicke} and \ref{sec:Dickegraphs}, we showed that the Dicke tower generated by $\Sd$ on $\ket{\overline 0}$ has the property that if these states are eigenstates of a local Hamiltonian, the states must be equally spaced in energy.
In this section, we will first introduce a theorem that takes generalized versions of the propositions \textit{as assumptions}.
We will then show that these assumptions hold for states generated by a wide class of ``quasiparticle creation operators" $\Qd$, and hence that those respective towers must be equally spaced in energy as well. We will assume that the system is on a graph with $N$ sites and maximum degree $\Delta$.

To keep the presentation simple, we will be restricting to states $\ket{Q^p} \propto (\Qd)^p\ket{\overline 0}$, where $\Qd$ has a definite charge under commutation with $\sum_i \sdag_i \s_i$: $[\sum_i \sdag_i \s_i, \Qd]  = \charge \Qd$. Then $\sum_i \sdag_i \s_i \ket{Q^p} = \charge p \ket{Q^p}$.

\begin{theorem}[Generalized Towers]\label{th:generalized}~\\
    Given a quasiparticle creation operator $\Qd$, if the states $\ket{Q^p} \equiv (\Qd)^p\ket{\overline 0}$ satisfy the following properties, where $\alpha, \beta, \gamma$ are each finite-valued functions:
    \begin{enumerate}
        \item[$P.1$] \label{Q1} For a range $R$ Hamiltonian $H$ satisfying $H\ket{Q}=E\ket{Q}$, $H$ can be written as \begin{equation}\label{eq:Qdecomp}H= \Omega' I + \omega' \sum_i \sdag_i \s_i + \sum_{X:\diam{X}\leq \alpha(R)} h_X\end{equation}
        for some constants $\Omega', \omega'$ and where $h_X \ket{Q}=h_X\ket{\overline{0}} =0$ and $h_X$ is supported entirely within $X$. 
        \item[$P.2$] \label{Q2} Given a maximum range $\pRmax$ and a collection of operators $h_X$ supported on $X$, where $h_X\ket{Q}= h_X \ket{\overline{0}} = 0$, if \begin{equation}
        \left( \sum_{X:\diam(X)\leq \pRmax} h_X \right) \ket{Q^p} = E'_p\ket{Q^p}
        \end{equation}
        for some energies $E'_p$, then $E'_p=0$ for $p\leq N/\beta(\pRmax)$. 
        \item[$P.3$] \label{Q3} Given a $k$-local Hamiltonian $H$, if 
    $H \ket{W^q} = 0$ for $q \leq \gamma(k)$, then $H \ket{Q^p}=0$ for all $p$.
    \end{enumerate}
Then for a range $R$, $k$-local Hamiltonian $H$ with $H\ket{Q^p} = E_p\ket{Q^p}$ and system size $N > \beta(\alpha(R)) \gamma(k)$, $E_p = \Omega + \omega p$ for constants $\Omega=\Omega',\omega=c\omega'$. 
\end{theorem}
\textit{Proof.}
The proof follows nearly identically to that of Theorem~\ref{th:dicke}; we include the following for completeness. 
Our assumptions are $H$ is a range $R$, $k$-local Hamiltonian with $H\ket{Q^p}=E_p\ket{Q^p}$ and system size $N>\beta(\alpha(R)) \gamma(k)$.

Since $H$ has $\ket{Q}$ as an eigenstate, by property $P.1$, $H$ can be written as Eq.~(\ref{eq:Qdecomp}).  Applying this form to $H \ket{Q^p} = E_p \ket{Q^p}$ and rearranging shows that

\begin{equation}
    \left( \sum_{X:\diam(X)\leq \alpha(R)} h_X \right) \ket{Q^p} = [E_p-(\Omega' + c\omega' p)]\ket{Q^p} \equiv E_p'\ket{Q^p} ~.
\end{equation}

If we can show that in fact $E_p' = 0$, we will have proven the theorem for $\Omega = \Omega'$ and $\omega = \charge \omega'$. 
The proof that $E_p'=0$ proceeds in two steps. 
First, we can immediately apply the assumed property $P.2$ with $\pRmax = \alpha(R)$.
Thus $E_p' = 0$ for $p \leq N/\beta(\alpha(R))$.
Since $N>\beta(\alpha(R))\gamma(k)$, then $E_p'=0$ for at least $p=0,1,...,\gamma(k)$.  Thus, the assumed property $P.3$ shows $E_p' = 0$ for all $p$. \hfill $\blacksquare$

In Propositions~\ref{prop:Q1class},~\ref{prop:Q2class}, and~\ref{prop:Qcommutator} that we will present below, we introduce classes of $\Qd$ for which the properties $P.1, P.2, P.3$ respectively hold. Prop.~\ref{prop:Qcommutator} concerns a class of $\Qd$ strictly larger than that of Prop.~\ref{prop:Q2class}, and Prop.~\ref{prop:Q2class} concerns a class of $\Qd$ strictly larger than that of Prop.~\ref{prop:Q1class}. The class of $\Qd$ in Prop.~\ref{prop:Q1class} thus satisfies all of the assumptions of Theorem~\ref{th:generalized}.

We give detailed discussions of these propositions and their proofs below, but we will first highlight some of their consequences. In particular, we emphasize a couple simple and characteristic choices of $\Qd$ that by virtue of Propositions~\ref{prop:Q1class},~\ref{prop:Q2class}, and~\ref{prop:Qcommutator} enjoy the properties of $P.1,P.2,P.3$ for some choices of $\alpha,\beta,\gamma$.

\begin{corollary}
    Define the 1D quasiparticle creation operator $\doub \equiv \sum_i \sdag_i \sdag_{i+1}$. As a corollary of Propositions~\ref{prop:Q1class},~\ref{prop:Q2class}, and~\ref{prop:Qcommutator}, $\doub$ satisfies the assumptions of Theorem~\ref{th:generalized}. 
\end{corollary}

\begin{corollary}
    On regular graphs of finite degree, define $\Qd_{\rm n.n.} \equiv \sum_i \prod_{j \in B_i(1)} \sdag_j$. $\rm n.n.$ stands for nearest neighbor; this quasiparticle creation operator consists of raising terms for each site and its nearest neighbors. As a corollary of Propositions~\ref{prop:Q1class},~\ref{prop:Q2class}, and~\ref{prop:Qcommutator}, $\Qd_{\rm n.n.}$ satisfies the assumptions of Theorem~\ref{th:generalized}. 
\end{corollary}
\textit{Comment.} Quasiparticle creation operators are often associated with some simple algebraic structure, such as a low-rank Lie algebra~\cite{odea2020from, ren2021quasisymmetry, ren2022deformed}. For example, in the Dicke tower context, $\mathfrak{su}(2)$ is generated when treating $\Sd$ and $S$ as the generators of a Lie algebra whose Lie bracket is the commutator. 
However, neither $\doub$ nor $\Qd_{\rm n.n.}$ generate simple Lie algebras under commutators with their respective Hermitian conjugates. 
Nevertheless, our results show that they generate towers on $\ket{\overline 0}$ that have strikingly similar properties to the Dicke tower: under local Hamiltonians, such towers must be equally spaced in energy, and the spacing can in fact be induced by $\sum_i \sdag_i\s_i$, just like the Dicke tower. 
\subsection{Properties of the generalized quasiparticle creation operator}
\subsubsection{\textit{P.1: Decomposition of $H$}}
Property $P.1$ is only concerned with states $\ket{\overline 0}$ and $\ket{Q}$. In particular, it is guaranteed to hold whenever $\ket{Q} = M \ket{W}$ and $\ket{\overline{0}} = M \ket{\overline{0}}$ for a locality-preserving map $M$.

In general, we will call a map $M$ $\delta$-locality-preserving if it is invertible and for every range $R$ operator $O$, $MOM^{-1}$ and $M^{-1} O M$ are at most range $R+\delta$ operators for some $\delta$ that is independent of system size. 

\begin{proposition}[$P.1$ from locality-preserving maps]\label{prop:locality}
    Let $M$ be a $\delta$-locality-preserving map.
    If $\ket{Q} = M \ket{W}$ and $\ket{\overline{0}} = M \ket{\overline{0}}$, then:
    for a range $R$ Hamiltonian $H$ satisfying $H\ket{Q}=E_1\ket{Q}$ and $H\ket{\overline 0}=E_0 \ket{\overline{0}}$, $H$ can be written as \begin{equation}\label{eq:QdecompQ1}H= \Omega I + \omega \sum_i \sdag_i \s_i + \sum_{X:\diam(X)\leq \alpha(R)} h_X\end{equation}
        for some constants $\Omega, \omega$, and where $h_X \ket{Q}=h_X\ket{\overline{0}} =0$ and $h_X$ is supported entirely within $X$. Explicitly, we bound $\alpha(R) \leq \Rmax +3\delta$.
\end{proposition}

\textit{Proof.}
By assumption, the operator $H'$ defined as
\begin{equation}
H'=H-[E_0 I + (E_1 - E_0) \sum_i \sdag_i \s_i]
\label{eq:Hpdefn}
\end{equation}
satisfies $H'\ket{Q}=H'\ket{\overline 0}=0$. As a consequence, $M^{-1} H'M\ket{W}= M^{-1} H'M\ket{\overline 0} = 0$. But $M^{-1} H' M$ has range at most $R+\delta$, so by Prop.~\ref{prop:decomp}, 
\begin{equation}
    M^{-1} H' M = \sum_{X':\diam(X')\leq 2(R+\delta)} \tilde{h}_{X'} ~,
\end{equation}
with $\tilde{h}_{X'}\ket{W}=\tilde{h}_{X'}\ket{\overline 0}=0$.

Transforming back, we have
\begin{equation}
    H' = \sum_{X':\diam(X')\leq 2(R+\delta)} M\tilde{h}_{X'}M^{-1} ~,
\label{eq:Hpsum}
\end{equation}
where we note that $M \tilde{h}_{X'} M^{-1}$ has range at most $2(R+\delta)+\delta = 2R +3\delta$. Furthermore, it satisfies $M \tilde{h}_{X'} M^{-1} \ket{\overline 0}=M\tilde{h}_{X'} M^{-1} \ket{Q}=0$.
Repackaging the sum over $X'$ in Eq.~(\ref{eq:Hpsum}) as a sum over the new supports $\{X\}$ of $\{M\tilde{h}_{X'} M^{-1} \}$ and using Eq.~(\ref{eq:Hpdefn}) to express $H$ thus completes the proof.\hfill $\blacksquare$

Useful choices of $M$ can often be constructed through a circuit geometry (i.e. $M$ can be broken into layers composed of disjoint gates). Before proceeding, we note the following well-known geometrical lemma useful for understanding the number of layers needed for a circuit in a general graph geometry. The proof is included for completeness in Appendix~\ref{app:geomproofs}.

\begin{lemma}[Disjoint layers]\label{lemma:layers}
On a graph with $N$ vertices and maximum vertex degree $\Delta$, consider the set of $N$ distinct balls of radius $r$ centered on each of the sites $i$, $\{B_i(r)\}$. It is possible to partition this set into $\Delta^{2r}+1$ disjoint subsets, each with the property that the balls in any given subset are themselves disjoint from one another.
\end{lemma}
Lemma~\ref{lemma:layers} is useful in the context of mapping between states using quantum circuits, particularly in establishing upper bounds on the number of layers needed to apply a set of $N$ quantum gates. For our purposes in Prop.~\ref{prop:Q1class}, the order that the gates are applied does not matter, but the gates in a given layer must be disjoint from one another. If the support of the $i$th gate is inside $B_i(r)$ for all $i$, then Lemma~\ref{lemma:layers} demonstrates that the $N$ quantum gates can be split into at most $\Delta^{2r}+1$ layers composed of gates with disjoint support. 
This will be useful in Prop.~\ref{prop:Q1class} when we need to establish bounds on  $\delta$ for a given $M$; having a finite number of layers consisting of disjoint gates of finite support will then lead to a finite lightcone.
In the following, we will describe a class of multibody quasiparticle creation operators for which it is easy to show the existence of a locality-preserving $M$. To lighten the notation, we will denote $\ket{0}_J \equiv \otimes_{j \in J} \ket{0}_j$ for some indexing set $J$.

\begin{proposition}\label{prop:Q1class}
Let $\Qd$ take the form $\sum_i \Qd_i$ where $\Qd_i = \prod_{j \in I_i} \sdag_j$ with index sets $I_i$. For each $i$, $I_i$ is a set of indices satisfying three conditions
\begin{enumerate}
    \item $|I_i|=\charge>1$;
    \item $|\mathbf{r}_j-\mathbf{r}_i| \leq d$ for all $j \in I_i$ for some constant $d$;
    \item If $i_1 \neq i_2$, then $I_{i_1} \neq I_{i_2}$.  To clarify, $I_{i_1}$ and $I_{i_2}$ for $i_1 \neq i_2$ may overlap, but they cannot be equal as sets.
\end{enumerate}
Then there exists a $\delta$-locality-preserving $M$ such that $\Qd\ket{\overline 0} = M \ket{W}$ and $\ket{\overline{0}} = M \ket{\overline{0}}$. Explicitly, on a graph with maximum degree $\Delta$, we bound $\delta \leq 8d(\Delta^{4d}+1)$.

As a consequence, by Prop.~\ref{prop:locality}, such $\Qd$ satisfy $P.1$ with $\alpha(R)\leq 2R+3\delta \leq 2R+24d(\Delta^{4d}+1).$
\end{proposition}
\textit{Comment.} 
This choice of $\Qd$ is such that there exists a bijection from each site $i$ to $\Qd_i$, with $\Qd_i$ close to $i$. The choice of the indexing sets is often not unique; for example, for $\Qd = \sum_i s^\dagger_i s^\dagger_{i+1}$ the choices $I_i=\{i-1,i\}$ and $I_i = \{i,i+1\}$ both correspond to $\charge = 2, d=1$. 
In particular, the bound involving $d$ gives a useful way to both bound the range of terms entering $\Qd$ and to ensure that they are each close to a lattice site. As an example, on the $2D$ square lattice, $\Qd = \sum_{i} \sdag_i \prod_{j:j\sim i} \sdag_j$ (where $j \sim i$ means that $j,i$ are nearest neighbors) satisfies the conditions of Prop.~\ref{prop:Q1class} with $\charge =5$, $d=1$.

\textit{Proof.} 
A valid choice of $M$ can be constructed through the following algorithm. For each site $i$, consider the ball $\bi \equiv B_i(2d)$ surrounding $i$. Define $P_i$ as the projector onto the states $\Qd_i\ket{0}_{\bi}$ and $\sdag_i\ket{0}_{\bi}$. Then we define \begin{equation}
    M_i = (I-P_i)+\left(\Qd_i\ket{0}_{\bi}\bra{0}_{\bi}\s_i + \sdag_i \ket{0}_{\bi} \bra{0}_{\bi} Q_i\right)
\end{equation}
$M_i$ will serve as an invertible (in fact, $M_i^{-1} = M_i$) building block of $M$; it converts the term $\sdag_i \ket{\overline 0}$ in $\ket{W}$ to the term $\Qd_i \ket{\overline 0}$ in $\ket{Q}$ and vice-versa, while keeping all other states invariant.

$M_i$ also enjoys the useful property that for $i \neq j$, $M_i M_j \sdag_j \ket{\overline 0} = M_j \sdag_j\ket{\overline 0} = Q_j^\dagger \ket{\overline 0}$. This is a direct consequence of the fact that  $Q_j^\dagger \ket{\overline 0}$ is orthogonal to both $\sdag_i \ket{0}_{\bi}$ and $\Qd_i\ket{0}_{\bi}$, which are the only two states on which $M_i$ acts non-trivially.
The orthogonality with $\sdag_i \ket{0}_{\bi}$ is a consequence of $\bi$ being a ball of radius $2d$:  $Q_j^\dagger \ket{\overline 0}$ either does not contain a particle at site $i$, or there is at least one additional particle within the radius $2d$ (conditions 1 and 2). Condition 3 ensures that the term $\sdag_i \ket{0}_{\bi} \bra{0}_{\bi} Q_i$ vanishes on  $Q_j^\dagger \ket{\overline 0}$. 

This property ensures that the order that the $M_i$ are applied to $\ket{W}$ does not matter; $\prod_j M_j \ket{W} = \frac{1}{\sqrt{N}} \sum_i Q_i^\dagger \ket{\overline 0} = \ket{Q}$ for any order of applying the $M_j$ gates in $\prod_j M_j$. 

The order in which the $M_i$ gates are applied can be chosen to ensure that the resulting $M$ is locality-preserving for some $\delta$. In particular, we can apply a quantum circuit built out of layers, with each layer consisting of a product of $M_i$ with disjoint support. The number of such layers, by Lemma~\ref{lemma:layers}, is at most $\Delta^{4d}+1$. Furthermore, because each $M_i$ has finite range $4d$,  a layer can only increase the range of an operator by at most $8d$. Thus,  the resulting circuit can only increase the range of an operator by at most $8d(\Delta^{4d}+1)$. \hfill $\blacksquare$

We will use the 1D quasiparticle creation operator $\doub= \sum_i \sdag_i \sdag_{i+1}$ to illustrate this construction in a simple way. To keep the construction at its simplest, we will take the system size $N$ to be a multiple of $5$. The choice $\Qd_i = \sdag_i \sdag_{i+1}$ satisfies the conditions of Prop.~\ref{prop:Q1class} with $c=2,d=1$. Then $P_i$ and $M_i$ take the form
\begin{equation}
P_i = \ket{00110}\bra{00110} + \ket{00100}\bra{00100}
\end{equation}
\begin{equation}
    M_i = (1-P_i) +(\ket{00110}\bra{00100}+\ket{00100}\bra{00110})
\end{equation}
with the middle spin being $i$. Our construction merely gives an upper bound (of 17) on the number of layers then needed to construct $M$, but the explicit choice of $M=L_5 L_4L_3L_2L_1$ with $L_j = \otimes_{i=0}^{N/5-1} M_{5i+j}$ shows that only five layers are needed. 

We note in passing that the particular choice of gates detailed in Prop.~\ref{prop:Q1class} also does not necessarily give the lowest range $M$; in the $\doub$ example, (taking $N$ a multiple of $3$ for simplicity) the choice of three-body gates
\begin{equation}
\tilde{P}_i = \ket{011}\bra{011} + \ket{010}\bra{010}
\end{equation}
\begin{equation}
    \tilde{M}_i = (1-\tilde{P}_i) +(\ket{011}\bra{010}+\ket{010}\bra{011})
\end{equation}
and $\tilde{M}=\tilde{L}_3 \tilde{L}_2 \tilde{L}_1$ with $\tilde{L}_j = \otimes_{i=0}^{N/3-1} \tilde{M}_{3i+j}$ works just as well. Our bound of $\delta \leq 136$ is thus quite loose; this last construction involves three layers consisting of products of three site gates, which is $\delta$-locality-preserving with $\delta= 18$.

Finally, we highlight that $M$ in Prop.~\ref{prop:locality} only needs to be invertible and locality-preserving (rather than unitary). The class of operators $\Qd$ considered in Prop.~\ref{prop:Q1class} can then be made slightly wider. In particular, for $\Qd_i$ satisfying the conditions of Prop.~\ref{prop:Q1class}, $\Qd = \sum_i a_i \Qd_i$ with $a_i \neq 0$ also works, with the choice of 
\begin{equation}
    M_i = (I-P_i)+\left(a_i \Qd_i\ket{0}_{\bi}\bra{0}_{\bi}\s_i + \frac{1}{a_i}\sdag_i \ket{0}_{\bi} \bra{0}_{\bi} Q_i\right)
\end{equation}
This $M_i$ is invertible (it is again its own inverse) and gives rise to an $M$ that is $\delta$-locality-preserving with the same bound on $\delta$ as in the proof of Prop.~\ref{prop:Q1class}.

\subsubsection{\textit{P.2: Annihilation of finite fraction}}
We will demonstrate that property $P.2$ holds for a somewhat wider class of $\Qd$ than considered in the previous subsection. 

\begin{proposition}[P.2]\label{prop:Q2class}
Let $\Qd$ take the form $\sum_{Y \in \mathcal{Y}} \Qd_Y$, where $\Qd_Y = \prod_{j \in Y} \sdag_j$ is labeled by its support $Y$. $\mathcal{Y}$ is a collection of subsets of indices satisfying the following two properties for some fixed constants $d_1, d_2$. Every $Y \in \mathcal{Y}$ satisfies $|Y|=c$ and $\diam(Y)\leq d_1$. Furthermore, for every site $i$, there exists some $Y^{(i)} \in \mathcal{Y}$ for which at least one site in $Y^{(i)}$ is within distance $d_2$ of site $i$.

Then given a maximum range $\pRmax$ and a collection of operators $h_X$ supported on $X$, where $h_X\ket{Q}= h_X \ket{\overline{0}} = 0$, if \begin{equation}
        \left( \sum_{X:\diam(X)\leq \pRmax} h_X \right) \ket{Q^p} = E'_p\ket{Q^p},
        \end{equation}
for some energies $E'_p$, then $E'_p=0$ for $p\leq N/\beta(\pRmax)$.

As a consequence, such $\Qd$ satisfy condition $P.2$; explicitly, on a graph of maximum degree $\Delta$, we bound $\beta(\pRmax) \leq \Delta^{2(d_1-1)+2d_2+\pRmax}+1$.
\end{proposition}
\textit{Proof.}  By assumption, $h_X \ket{Q}=h_X \ket{\overline 0} =0$. It will also be useful to note that
\begin{equation}
\begin{split}\label{eq:Qann}
h_X \sum_{Y \in \mathcal{Y}, Y\cap X \neq \emptyset} \Qd_Y \ket{\overline 0} &=  h_X \left( \sum_{Y \in \mathcal{Y}} \Qd_Y \ket{\overline 0} -\sum_{Y \in \mathcal{Y}, Y\cap X = \emptyset} \Qd_Y \ket{\overline 0} \right) \\&= h_X \ket{Q}-\sum_{Y \in \mathcal{Y}, Y\cap X = \emptyset}\Qd_Y h_X\ket{\overline 0} = 0
\end{split}
\end{equation}

We introduce the normalization constants $\mathcal{N}_p$ (as in $\ket{Q^p} = \mathcal{N}_p (\Qd)^p \ket{\overline 0}$); we will not need any special properties of the $\mathcal{N}_p$.
As in the case of the Dicke tower's Prop.~\ref{prop:densityann}, we will build intuition by discussing the case of $\ket{Q^2} \propto (\Qd)^2 \ket{\overline 0}$ first.  Through $h_X\ket{\overline 0}=0$ and Eq.~\eqref{eq:Qann},
\begin{equation}
\begin{split}
    &\frac{1}{\mathcal{N}_2} h_X \ket{Q^2} = h_X \sum_{Y_2 \in \mathcal{Y}}\sum_{Y_1 \in \mathcal{Y}} \Qd_{Y_2} \Qd_{Y_1} \ket{\overline 0}
    \\&= h_X \sum_{\substack{Y_2 \in \mathcal{Y} \\ Y_2 \cap X\neq \emptyset}} \sum_{\substack{Y_1 \in \mathcal{Y} \\ Y_1 \cap X\neq \emptyset}} \Qd_{Y_2} \Qd_{Y_1} \ket{\overline 0}
    +2h_X \sum_{\substack{Y_2 \in \mathcal{Y} \\ Y_2 \cap X= \emptyset}} \sum_{\substack{Y_1 \in \mathcal{Y} \\ Y_1 \cap X\neq \emptyset}} \Qd_{Y_2} \Qd_{Y_1} \ket{\overline 0} 
    +h_X \sum_{\substack{Y_2 \in \mathcal{Y} \\ Y_2 \cap X= \emptyset}} \sum_{\substack{Y_1 \in \mathcal{Y} \\ Y_1 \cap X= \emptyset}} \Qd_{Y_2} \Qd_{Y_1} \ket{\overline 0}
    \\&= h_X \sum_{\substack{Y_2 \in \mathcal{Y} \\ Y_2 \cap X\neq \emptyset}} \sum_{\substack{Y_1 \in \mathcal{Y} \\ Y_1 \cap X\neq \emptyset}} \Qd_{Y_2} \Qd_{Y_1} \ket{\overline 0}
    +2 \sum_{\substack{Y_2 \in \mathcal{Y} \\ Y_2 \cap X= \emptyset}} \Qd_{Y_2}h_X\sum_{\substack{Y_1 \in \mathcal{Y} \\ Y_1 \cap X\neq \emptyset}}  \Qd_{Y_1} \ket{\overline 0} 
    + \sum_{\substack{Y_2 \in \mathcal{Y} \\ Y_2 \cap X= \emptyset}} \sum_{\substack{Y_1 \in \mathcal{Y} \\ Y_1 \cap X= \emptyset}} \Qd_{Y_2} \Qd_{Y_1} h_X\ket{\overline 0}
    \\&= h_X \sum_{\substack{Y_2 \in \mathcal{Y} \\ Y_2 \cap X\neq \emptyset}} \sum_{\substack{Y_1 \in \mathcal{Y} \\ Y_1 \cap X\neq \emptyset}} \Qd_{Y_2} \Qd_{Y_1} \ket{\overline 0} + 0 + 0
\end{split}
\end{equation}
Thus, for $\sum_{X: \diam(X) \leq \pRmax} h_X \ket{Q^2} = E_2' \ket{Q^2}$, we have
\begin{equation}\label{eq:Qsquaredstate}
    \mathcal{N}_2 \sum_{X: \diam(X) \leq \pRmax} h_X \sum_{\substack{Y_2 \in \mathcal{Y} \\ Y_2 \cap X\neq \emptyset}} \sum_{\substack{Y_1 \in \mathcal{Y} \\ Y_1 \cap X\neq \emptyset}} \Qd_{Y_2} \Qd_{Y_1} \ket{\overline 0} = E_2' \ket{Q^2}
\end{equation}
Crucially,  every basis state within the left hand side of this expression has $2\charge$ particles within some ball of radius $(d_1-1) +\pRmax/2$. 
To see this, note that for every $X,Y_1,Y_2$ in the sum, the $2\charge$ particles in $h_X \Qd_{Y_1} \Qd_{Y_2} \ket{\overline 0}$ must be within $X \cup Y_1 \cup Y_2$, and $Y_1$ and $Y_2$ both intersect $X$ nontrivially. 
Thus the farthest two sites can be within $X \cup Y_1 \cup Y_2$ is at most $(\diam(Y_1)-1) + (\diam(Y_2)-1) + (\diam(X)-1) \leq 2(d_1-1)+\pRmax-1$, so that there exists a ball of radius $(d_1-1)+\pRmax/2$ that covers the $2\charge$ particles.

On the other hand, when $E_2' \neq 0$ and $N$ is large enough, there are basis states within the right hand side of this expression for which there is no ball of radius $(d_1-1) +\pRmax/2$ that contains $2 c$ particles. 
This is possible so long as $N$ is large enough that the system can fit two disjoint balls of radius $(d_1-1) +\pRmax/2+d_2$. Call the centers of these two disjoint balls $i$ and $j$. Then by assumption there are $Y^{(i)}$ and $Y^{(j)}$ that are $d_2$-close to $i$ and $j$ respectively. 
Having chosen the radii of the disjoint balls to be $(d_1-1) +\pRmax/2 + d_2$, the smallest ball containing the support of both $Y^{(i)}$ and $Y^{(j)}$ must have a radius larger than $(d_1-1)+\pRmax/2$. 
Thus $\Qd_{Y^{(i)}}\Qd_{Y^{(j)}}\ket{\overline 0}$ is a basis state in $\ket{Q^2}$ that is not in the left hand side of Eq.~\ref{eq:Qsquaredstate}, and the only way this is possible is if $E_2'=0$. 

The proof proceeds similarly for $\ket{Q^p}$. For any $p$, a ball of radius $(d_1-1) +\pRmax/2$ can cover $\left(\bigcup_{a=1}^p Y_a\right)\cup X $ with $Y_a \in \mathcal{Y}$ so long as $Y_a \cap X \neq \emptyset$ for all $a \in \{1,..,p\}$. If $\sum_{X: \diam(X) \leq \pRmax} h_X \ket{Q^p} = E_p' \ket{Q^p}$, then for each basis state in $\sum_{X: \diam(X) \leq \pRmax} h_X \ket{Q^p}$ there must be a ball of radius $(d_1-1)+\pRmax/2$ that contains at least $2\charge$ particles. However, if the graph contains $p$ disjoint balls of radius $(d_1-1) +\pRmax/2 + d_2$, there will exist a basis state in $\ket{Q^p}$ that is not present in $\sum_{X: \diam(X) \leq \pRmax} h_X \ket{Q^p}$ (i.e. one for which there is no ball of radius $(d_1-1)+\pRmax/2$ that contains at least $2\charge$ particles). This forces $E_p'=0$. By Lemma~\ref{lemma:sphere}, such a sphere packing is possible so long as $p<N/(\Delta^{2(d_1-1)+\pRmax+2d_2}+1)$. \hfill $\blacksquare$

\subsubsection{\textit{P.3: Induction on annihilation}}
We can demonstrate property $P.3$ for an even wider class of quasiparticle creation operators $\Qd$ that subsumes those discussed in the previous two subsections.

We will in fact show $P.3$ with $\gamma(k) = 2k$ for every quasiparticle creation operator that is a linear combination of products of $\sdag_i$. 
Our argument will also not rely on $\Qd$ having a definite charge $\charge$ under $\sum_i \sdag_i \s_i$, so we will relax this condition for this subsection only. 
This is a very general class of quasiparticle creation operators; to give a sense of its generality, we note that it includes the 1D $\Qd = \sum_i \sdag_i - \sum_i\sdag_{i} \sdag_{i+1}\sdag_{i+2}$, and it includes $\Qd = \sum_{i} \sdag_i \prod_{j: j \sim i} \sdag_j$ on arbitrary bounded-degree graphs. 

Property $P.3$ is directly analogous to the Dicke tower's Prop.~\ref{prop:annall}, which itself was a consequence of iterated commutators of $\Sd$ vanishing, as in Lemma~\ref{lemma:commutator}. Indeed, if $\Sd \to \Qd$ and $2k \to \gamma(k)$ in the proof of Prop.~\ref{prop:annall}, we see that

\centerline{``For $k$-local $O$, $[[[[O,\Qd],\Qd],...],\Qd]_{\gamma(k)+1} = 0$"}

\centerline{$\implies$ ``For $k$-local $H$, if $H\ket{Q^p}=0$ for $p \leq \gamma(k)$, then $H\ket{Q^p}=0$ for all $p$."}

Thus the key property to show is the following proposition.
\begin{proposition}[Nilpotent commutator]\label{prop:Qcommutator}~\\
    Let $\Qd$ be a linear combination of products of $\sdag_i$. For any $k$-local $O$, $[[[[O,\Qd],\Qd],...],\Qd]_{2k+1}=0$. Thus such $\Qd$ satisfy property $P.3$ with $\gamma(k) = 2k+1$.
\end{proposition}
\textit{Proof.}
The proof is analogous to that of Lemma~\ref{lemma:commutator}, with one small change.

First, as a characteristic example, consider the 1D multibody quasiparticle creation operator $\doub =\sum_i \sdag_i \sdag_{i+1}$. 
As we noted  in the proof of Lemma.~\ref{lemma:commutator}, the support of a commutator $[\Sd,O]$ will be equal to or within the support of $O$. However, the support of a commutator $[\doub,A]$ could have support on \textit{a larger region} than $A$. For example, $[\doub,\s_j] = 2 \sdag_{j-1} s^z_j + 2s_j^z \sdag_{j+1}$; the commutators of $\sdag_{j-1} \sdag_{j}$ and $\sdag_j \sdag_{j+1}$ with $\s_j$ are nontrivial only at site $j$ and leave behind excess $\sdag$ whose support did not touch that of $\s_j$. 

This ``excess support" will occur quite generally for the $\Qd$ of interest. However, because $\Qd$ consists only of linear combinations of products of $\sdag_i$, the excess support generated will either be empty or consist entirely of factors of $\sdag_i$ and not $s_i$; such factors cannot ``anchor" further iterated commutators with $\doub$, as $[\sdag_i,\sdag_i]=0$ trivially. 
Thus a $(2k+1)$-fold iterated commutator of $\doub$ with any $k$-local $O$ will still annihilate $O$, even if an intermediate number of commutators with $O$ is no longer $k$-local for the same $k$.

An alternative argument is to note that a $k$-local $O$ can at most lower the restriction of total $S^z$ on its support $\text{supp}(O)$ by $k$, while each commutation with $Q^\dagger$ raises it by at least $1$,  and after $2k+1$ commutations the result would annihilate any state.
\hfill $\blacksquare$

We wrap up the discussion of property $P.3$ with a few comments.

For the sake of showing Prop.~\ref{prop:Qcommutator}, $\Qd$ itself can be very nonlocal; even if $\Qd$ is $k'$-local with $k'\gg k$, $\gamma(k)$ remains $2k$. However, in demonstrating the properties $P.1$ and $P.2$ for families of $\Qd$, locality of the considered $\Qd$ played a more important role. 

If $\ell$ bosons could fit per site, rather than just one, then for $\Qd$ consisting of products of $\sdag$, the corresponding $\gamma(k)=2\ell k$. In the spin language, with spin-$S$ particles on each site, this would correspond to $\gamma(k)=4 S k$. However, we will not consider such a generalization further.

Quasiparticle creation operators that are sums of terms that \textit{do not} mutually commute will generically \textit{not} give rise to nilpotent iterated commutators with a small $\gamma(k)$. For example, an iterated commutator of $\Qd = \sum_i \sdag_i \sdag_{i+1} \s_{i+2}$ with a $k$-local $O$ need not vanish until the number of iterated commutators is comparable to system size, even for small $k$. We also do not consider such cases in this work.

To conclude our discussion of families of $\Qd$ that enjoy the three properties $P.1,P.2,P.3$, we informally revisit our assumptions on $\Qd$. We have shown $P.3$ holds for $\Qd$ equal to linear combinations of products of $\sdag_i$. For $P.2$, we additionally assume that $\Qd$ contains terms near every site, as this allows far apart particles in $\ket{Q^p}$. For $P.1$, in addition to the previous assumptions, we also require that the terms $\Qd_i$ of $\Qd$ generate orthogonal states; this allows us to construct a simple invertible mapping between $\ket{Q}$ and $\ket{W}$. We use the most restrictive assumptions for realizing $P.1$; as discussed above, these assumptions are nevertheless satisfied by many natural families of quasiparticle creation operators. 

\section{Discussion}\label{sec:discussion}

Our work demonstrates a locality-enforced strict ``spectral rigidity" of many quasiparticle towers of quantum many-body scars: when the states in one of these towers are eigenstates of an extensive local Hamiltonian, they necessarily come equally spaced in energy.
Under such Hamiltonian dynamics, superpositions of these states are then perfectly periodic in time, with period set by the energy difference between two consecutive states in the tower.
Our work lends itself to several interesting consequences and extensions, and we highlight several related avenues to explore.

\subsection{Entanglement freezing}
In Ref.~\cite{odea2025entanglement}, the following theorem was proven:
\begin{theorem}\label{th:freeze}
Given a Hamiltonian $H$, suppose there exists a sum of single-site operators that ``reproduces the energies" of a subset of eigenstates $\{\ket{\phi_n}\}$ of $H$, i.e. 
\begin{equation}
    \exists H_1 = \sum\nolimits_j h_j \text{ s.t. } H \ket{\phi_n} =H_1\ket{\phi_n} =  E_n \ket{\phi_n}.
\end{equation}
Here $h_j$ is an operator with support only at site $j$. Then all superpositions $\sum_n c_n \ket{\phi_n}$ of this subset of states will have time-independent entanglement under the dynamics generated by $H$.
\end{theorem}
In that work, the theorem was applied to several families of spin chain Hamiltonians with equally spaced scar towers. For many of these spin chain Hamiltonians, $H$ satisfied the condition of the theorem for the choice $H_1 \propto \sum_j S^z_j$.

The equal spacing theorems in the current work interface nicely with the entanglement freezing theorem.
For example, consider the Dicke tower of states.
The Hamiltonian $H_1=\Omega I + \omega \sum_{i} \sdag_i \s_i$ splits the Dicke tower equally in energy.
Then by Theorem~\ref{th:freeze}, any Hamiltonian that splits these states equally in energy cannot generate entanglement dynamics on any superposition of the Dicke tower states. 
However, by Theorems~\ref{th:dicke} and~\ref{th:dickegraphs}, we know that local Hamiltonians with the Dicke tower as eigenstates will \textit{necessarily} split those states equally in energy.
Thus, local Hamiltonians with the Dicke tower as eigenstates can \textit{never} generate entanglement dynamics starting in any superposition of the Dicke tower states.

Similarly, our generalized towers in Sec.~\ref{sec:general} focused on quasiparticle towers for which $H_1=\Omega I + \omega \sum_{i} \sdag_i \s_i$ splits the energies in the tower equally.
Accordingly, for those raising operators satisfying the assumptions of Theorem~\ref{th:generalized}, superpositions of the states in the corresponding towers cannot have any entanglement dynamics when they are eigenstates of a local Hamiltonian.

\subsection{Deformed towers}\label{subsec:deformed}
We emphasize that the equal spacing property is preserved under locality-preserving transformations,  and we will refer to towers obtained under such transformations as \textit{deformed towers}.
Suppose that the states $\ket{Q^p}$ are equally spaced whenever they are eigenstates of a local Hamiltonian. Define a deformed tower via $|\tilde{Q}^p\rangle = M\ket{Q^p}$ through an invertible and locality-preserving $M$. 
Then the states $|\tilde{Q}^p\rangle$ will also be equally spaced in energy whenever they are eigenstates of a local Hamiltonian.
This follows because any local Hamiltonian $H$ that has the $|\tilde{Q}^p\rangle$ states as eigenstates induces a local Hamiltonian $M^{-1}HM$ that has the $\ket{Q^p}$ as eigenstates with the same energies.  
Note that since our equal spacing theorems also hold for non-Hermitian parent Hamiltonians of the corresponding quasiparticle towers, $M$ need not be unitary.

We should emphasize that this method of deformation is related to but different from that considered in Prop.~\ref{prop:Q2class}.
There, we considered maps $M$ that only mapped the first two states of the quasiparticle tower to those of the Dicke tower. 
Throughout Sec.~\ref{sec:general}, we were also restricting to $\Qd$ with a fixed charge $c$ under $\sum_i \sdag_i \s_i$. In comparison, in this subsection, we are discussing maps from an entire quasiparticle tower to another entire quasiparticle tower.
Under a locality-preserving transformation, the operator that splits the energies equally will generically no longer be of the form $\sum_i \sdag_i \s_i$. It may no longer be a sum of single-site terms, either. In such a case, the deformed tower may show nontrivial entanglement dynamics (which is guaranteed to be periodic by the equal spacing of the energies).
However, as locality-preserving maps $M$ cannot greatly change the entanglement entropy of states, the amplitude of entanglement entropy oscillations will in turn be bounded~\cite{odea2025entanglement}.

\subsection{Relaxing locality}
In our equal spacing Theorems~\ref{th:dickegraphs} and~\ref{th:generalized}, we only needed a very weak condition on locality; the interactions need only be local with respect to a bounded degree graph.
Furthermore, as noted at the end of Sec.~\ref{sec:Dickegraphs}, any ``low-density" Hamiltonian where each term is $k$-local and each site is involved in at most $T_{\text{max} }$ terms induces a bounded-degree graph on which the Hamiltonian is geometrically local. While this is already a comparably weak constraint on locality, a natural question is whether the assumptions on locality can be further reduced. 

Mere $k$-locality of the Hamiltonian is not enough to guarantee equal spacings of the towers; for example, the Hamiltonian $\big(\sum_i \sdag_i \s_i \big)^2$ is 2-local but gives rise to a quadratic energy for the Dicke tower: $\big(\sum_i  \sdag_i \s_i \big)^2 \ket{W^p} = p^2\ket{W^p}$.
This example violates the ``low-density" condition, as each site is part of $N$ interactions, and hence cannot be expressed as a local Hamiltonian on some bounded-degree graph.

An alternative condition could be a version of quasilocality, such as range $R$ terms in $H$ being exponentially suppressed in $R$. In such a setting, the equal spacing might only be approximate, but the error relative to linear spacing might shrink as the system size grows. This setting is also natural when attempting to deform quasiparticle towers with mappings $M$ that only preserve quasilocality rather than strict locality; such kinds of mappings include adiabatic evolution, which can change the correlation lengths of states and map short-range entangled states to one another (see also the following subsection). We leave to future work the question of whether our theorems generalize to quasilocal Hamiltonians. 

\subsection{Towers built on quasiparticle vacua with non-zero correlation length}
For the Dicke tower, as well as the more general quasiparticle towers considered in Sec.~\ref{sec:general}, we restricted the base of the tower to be the vacuum of the hard-core bosons, $\ket{Q^0} = \ket{\overline 0}$.
However, in our discussion of deformed towers in Sec.~\ref{subsec:deformed}, the base of the deformed tower is $M\ket{\overline 0}$ for a locality preserving $M$, which does not need to equal $\ket{\overline 0}$.
On the other hand, a locality-preserving $M$ cannot induce a nonzero correlation length: correlations in $M\ket{\overline 0}$ will still be strictly zero beyond some $\Theta(1)$ distance.
An important future direction is to extend our results to quasiparticle towers for which the base of the tower, $\ket{Q^0}$, has a non-vanishing correlation length.
The quantum scar tower in the spin-$1$ AKLT model is a good example~\cite{moudgalya2018nonintegrable, moudgalya2018entanglement}; the scars are built on top of the AKLT ground state~\cite{AffleckKennedyLiebTasaki1987}, which has a non-zero correlation length, and all known local Hamiltonians split the scars equally in energy~\cite{Moudgalya2020Large, mark2020unified, odea2020from, moudgalya2023exhaustive}.
Furthermore, the AKLT ground state famously has a simple matrix product state representation~\cite{FannesNachtergaeleWerner1992, Perez-Garcia2007}, which may lend itself to a rigorous understanding of the scar states' energy spacing.
The AKLT ground state can be constructed by acting with projectors on a dimerized spin-$1/2$ state; this method of projecting is also useful for understanding the scars~\cite{moudgalya2018nonintegrable,moudgalya2018entanglement} and their parent Hamiltonians~\cite{keita2023projector2}.
However, the non-invertibility of the projections means that the discussion on deformed towers in Sec.~\ref{subsec:deformed} does not immediately apply. 
We leave the study of this conjecture of the equal spacing of the AKLT scar tower to future work.

\section{Acknowledgments}
We acknowledge support provided by the Princeton Center for Theoretical Science and the Princeton Quantum Initiative at Princeton University; by the Walter Burke Institute for Theoretical Physics at Caltech;
by the Institute for Quantum Information and Matter, an NSF Physics Frontiers Center (NSF Grant PHY-2317110); by the Munich Center for Quantum Science and Technology (MCQST) and the Deutsche Forschungsgemeinschaft (DFG, German Research Foundation) under Germany’s Excellence Strategy--EXC--2111--390814868; and by the National Science Foundation through grant DMR-2001186.
Nicholas O'Dea thanks Adithya Sriram and Long-Hin Tang for useful conversations on related topics. 

\textit{A related work by Keita Omiya that also shows equal spacing of energies of quasiparticle towers will appear in the same arXiv posting.
The proof techniques in each work are different, and the kinds of quasiparticle towers handled complement each other.
The works agree where they overlap; both prove the equal spacing of the Dicke tower.
We thank Keita Omiya for coordinating posting and for useful discussions.}

%%%%%%%%%%%%%%%%%%%%%%%%%%%%%%%%%%%%%%%%%%%%%%%%%%%%%%%%%%%%%%%%%%%%%%%%%%%%%%%
\appendix
\section{Proofs of geometrical lemmas}
\label{app:geomproofs}
\textit{Proof of Lemma~\ref{lemma:sphere}.}
Consider an algorithm that deletes regions of a graph according to the following rule. Pick a vertex $v$ at random and delete the ball $B_v(2r)$ surrounding it, and then repeat this rule on the remaining vertices until the graph is empty. On the original graph, the set of picked vertices defines the centers of a disjoint set of balls of radius $r$. This implies that the number of steps until the rule has deleted all vertices gives a lower bound on the number of disjoint balls of radius $r$ that can be packed on the graph. 

On a graph with maximum vertex degree $\Delta$, a ball of radius $n$ can fit at most $\Delta^n+1$ spins. Since the algorithm deletes a ball of radius $2r$ each step, it will take at least $N/(\Delta^{2r}+1)$ steps to terminate. \hfill $\blacksquare$

\textit{Proof of Lemma~\ref{lemma:layers}.}
Two balls $B_i(r)$ and $B_j(r)$ are disjoint if their centers $i,j$ satisfy $|\mathbf{r}_i - \mathbf{r}_j|>2r$. 

Consider the related problem of coloring the vertices of the graph, where no color is repeated in any ball of radius $2r$. Each such coloring will induce a valid partition of the set $\{B_i(r)\}$: each color corresponds to one of the disjoint subsets, and if site $i$ has some given color in the vertex coloring, then $B_i(r)$ is an element of the corresponding subset. An upper bound on the number of colors needed will thus give an upper bound on the number of subsets needed for a valid partitioning.

In turn, such a coloring is itself induced by a \textit{proper} coloring on a new graph; a proper coloring is one in which no two vertices connected by an edge can be the same color. The new graph has the same vertices as before, but its edge set also includes edges between any sites $i$ and $j$ that are within graph distance $2r$ of one another; a proper coloring on such a graph will ensure that no two vertices within radius $2r$ on the original graph have the same color. 

Furthermore, a proper coloring only ever requires at most one more color than the maximum degree: a vertex has a number of neighbors equal to at most the maximum degree, so a vertex is surrounded by a number of different colors that is at most the maximum degree, and hence with one more color than the maximum degree a vertex can always be colored differently than its neighbors.
This new graph has maximum degree set by the number of spins in a ball of radius $2r$ (minus one), which is at most $\Delta^{2r}$, so the number of colors needed is at most $\Delta^{2r}+1$. \hfill $\blacksquare$

\section{Proof of Prop.~\ref{prop:decompgengraph}: Decomposition of parent Hamiltonian for the $W$ state on general graph}
\label{app:Hdecompgengraph}
{\it Proof.}
As mentioned in the main text, this proposition extends results in Sec.~III of Ref.~\cite{gioia2025distinct}
to arbitrary graphs.
For easy reference we reproduce Table I from Ref.~\cite{gioia2025distinct} here in Table~\ref{tab:reftabI}.
\begin{table}[h]
    \centering
    \begin{tabular}{c|c|c|c}
        Operators in $H$, Eq.~(\ref{eq:G_normal_ordered_strings}) & $n$ & $m$ & Conditions\\\hline
        $c^{j_1...j_n} s_{j_1}^\dag ...s_{j_n}^\dag$ & $\geq 1$ & 0 & $c^{j_1...j_n}=0$\\
        $c^{j_1...j_n}_{k_1...k_m} s_{j_1}^\dag ...s_{j_n}^\dag s_{k_1}...s_{k_m} $ & $\geq 0$ & $\geq 2$ & None\\
        $\sum_k c^{j_1...j_n}_{k} s_{j_1}^\dag ... s_{j_n}^\dag s_{k}$ & $\geq 2$ & $1$ & $\sum_k c^{j_1...j_n}_k=0$ \\
        $\sum_k c_{k}  s_{k} $ & $0$ & $1$ & $\sum_k c_k=0$\\
        $\sum_k c^j_{k}s_j^\dag s_k$ & 1 & 1 & $\sum_{k}c^j_{k}= \lambda$
    \end{tabular}
    \caption{
    Characterization of an extensive local operator $H$ satisfying $H \ket{W} = \lambda \ket{W}$, via conditions on the expansion coefficients in the specific operator basis employing the hard-core boson creation/annihilation operator language, Eq.~(\ref{eq:G_normal_ordered_strings}).
    $H$ does not need to be Hermitian.
   }
\label{tab:reftabI}
\end{table}

All statements below refer to the expansion of $H$ in the normal-ordered operator string basis as in Ref.~\cite{gioia2025distinct},
\begin{equation}
H = \sum c^{j_1, \dots, j_n}_{k_1, \dots, k_m} s_{j_1}^\dagger \dots s_{j_n}^\dagger s_{k_1} \dots s_{k_m} ~.
\label{eq:G_normal_ordered_strings}
\end{equation}
Here $j_1, \dots, j_n$ are distinct sites and $k_1, \dots, k_m$ are distinct sites, but there can be overlap between the two sets. The range restriction corresponds to $\diam(\{j_1, \dots, j_n, k_1, \dots, k_m \}) \leq R$.
The first row in Table~\ref{tab:reftabI} splits into two parts, $n \geq 2, m = 0$ and $n = 1, m = 0$, which need slightly different arguments.
In the case \underline{$n \geq 2, m = 0$}, for assumed $c^{j_1, \dots, j_n} s_{j_1}^\dagger \dots s_{j_n}^\dagger$ present, we find a site $i$ that is at least as far away as $R$ from any of the sites $j_1, \dots, j_n$ (it is sufficient to find a site at least as far away as $2R$ from \textit{one} of the sites $j_1,...,j_n$).
Recall that in our definition of the range $R$, the distance between two points in any interaction term in Eq.~(\ref{eq:G_normal_ordered_strings}) is strictly less than $R$, so the site $i$ cannot be in any interaction term with the sites $j_1, \dots, j_n$.
From here the argument is identical to Ref.~\cite{gioia2025distinct}:
Configuration with $n+1$ of 1's on sites $j_1,\dots,j_n,i$ is produced by this specific term and cannot be produced by any other term, hence $c^{j_1, \dots, j_n}$ must be zero.
In the case \underline{$n = 1, m = 0$}, for assumed $c^j s_j^\dagger$ present, we find a site $i$ that is at least a distance $R$ away from $j$, and then as in Ref.~\cite{gioia2025distinct} deduce condition $c^j + c^i = 0$.
We then find a site $i'$ that is at least a distance $R$ away from both $j$ or $i$, deducing $c^j + c^{i'} = 0$ and $c^i + c^{i'} = 0$, and hence $c^j = 0$.

The second row in Table~\ref{tab:reftabI} trivially says that all terms with \underline{$n \geq 0, m \geq 2$} annihilate $\ket{W}$ and can be declared as $h_X$ terms by themselves (and all locality properties are just inherited); hence we can remove such terms from further considerations.

The third row in Table~\ref{tab:reftabI} treats terms \underline{$n \geq 2, m = 1$}, and for fixed $j_1, \dots, j_n$ it combines the corresponding operator strings into an $h_X = \sum_k c^{j_1, \dots, j_n}_k s_{j_1}^\dagger \dots s_{j_n}^\dagger s_k$ that annihilates $\ket{W}$ since $\sum_k c^{j_1, \dots, j_n}_k = 0$, where the argument for this condition is identical to Ref.~\cite{gioia2025distinct}: this collects all terms in $H$ that can produce a state with $n$ particles located at $j_1, \dots, j_n$, while there is no such state in the R.H.S.\ of $H \ket{W} = \lambda \ket{W}$.
Hence we only need to describe locality properties of the obtained $h_X$ term in the more general graph context.
In particular, the diameter of the exhibited $h_X$ is upper-bounded by $2R$ using the triangle inequality (which is satisfied by the graph distance).
Indeed, the support of $h_X$ is the union of $j_1, \dots, j_n$ and of all $k$ that appear in the sum; however, for any such $k'$ and $k''$, $|\mathbf{r}_{k'} - \mathbf{r}_{j_1}| < R$ and $|\mathbf{r}_{k''} - \mathbf{r}_{j_1}| < R$, hence $|\mathbf{r}_{k'} - \mathbf{r}_{k''}| < 2R$.

The fourth row in Table~\ref{tab:reftabI} dealing with \underline{$n = 0, m = 1$} terms, $\sum_k c_k s_k$, can be treated as in Ref.~\cite{gioia2025distinct} by reorganizing into a sum of 2-local annihilators of $\ket{W}$. Define the operators $\tau_{ij} = s_i - s_j$ for every nearest-neighbor pair of sites $i$ and $j$. First, we note that $\sum_k c_k = 0$. Then, on a connected graph, $\sum_k c_k s_k$ can be re-written as a linear combination of the $\tau_{ij}$; this means $\sum_k c_k s_k$ can be re-written as a linear combination of range-$2$ operators that individually annihilate the Hamiltonian. In particular, pick a spanning tree of the graph; a spanning tree is a connected subgraph without loops that contains all vertices of the original graph. Then, starting from the leaves, we eliminate $c_k s_k$ by subtracting $c_k \tau_{k,k'}$ where $k'$ is connected to $k$ on the tree: $c_k \to c_k - c_k= 0, c_{k'} \to c_{k'} + c_k$; this preserves the condition $\sum_l c_l = 0$. 
This proceeds until a single site is left, and the amplitude on that site is zero. 

Finally, the \underline{$n = 1, m = 1$} ``hard-core boson hopping'' terms in the last row of Table~\ref{tab:reftabI} can be rewritten as two-site non-Hermitian annihilators plus on-site terms:
\begin{equation}
s_j^\dagger s_k = (s_j^\dagger s_k - n_j) + n_j ~,
\end{equation}
where $n_j := \sdag_j s_j$ and it is easy to see that $(s_j^\dagger s_k - n_j) \ket{W} = 0$.
(This is an even faster derivation than in Ref.~\cite{gioia2025distinct} that noted that pure imaginary hopping parent Hamiltonian can also be rewritten in terms of non-Hermitian annihilators.)
Hence we can extract such non-Hermitian local annihilators of the $W$ state, i.e., define $h_X = c^j_k (s_j^\dagger s_k - n_j)$, which have exactly the same locality properties as the original $c^j_k s_j^\dagger s_k$.
We are then left with $H' = \sum_j c^{\prime j}_j n_j$ that has $\ket{W}$ as an eigenstate.
For this to be true, all $c^{\prime j}_j$ must be the same, $c^{\prime j}_j = \lambda,\; \forall j$, and hence $H^\prime = \lambda \sum_j{s^\dagger_j s_j}$.

\bibliography{equalspacingbib}

@article{wampler2025absorbing,
   title={Absorbing state phase transitions beyond directed percolation in dissipative quantum state preparation},
   volume={7},
   ISSN={2643-1564},
   url={http://dx.doi.org/10.1103/6pg7-3pxf},
   DOI={10.1103/6pg7-3pxf},
   number={3},
   journal={Physical Review Research},
   publisher={American Physical Society (APS)},
   author={Wampler, Matthew and Cooper, Nigel R.},
   year={2025},
   month=jul }

@ARTICLE{mohapatra2025unraveling,
       author = {{Mohapatra}, Sashikanta and {Moudgalya}, Sanjay and {Balram}, Ajit C.},
        title = "{Unraveling additional quantum many-body scars of the spin-$1$ $XY$ model with Fock-space cages and commutant algebras}",
      journal = {arXiv e-prints},
         year = 2025,
        month = nov,
          eid = {arXiv:2511.14878},
        pages = {arXiv:2511.14878},
          doi = {10.48550/arXiv.2511.14878},
archivePrefix = {arXiv},
       eprint = {2511.14878},
 primaryClass = {cond-mat.str-el},
       adsurl = {https://ui.adsabs.harvard.edu/abs/2025arXiv251114878M}
}

@misc{marconi2025symmetricquantumstatesreview,
      title={Symmetric quantum states: a review of recent progress}, 
      author={Carlo Marconi and Guillem Müller-Rigat and Jordi Romero-Pallejà and Jordi Tura and Anna Sanpera},
      year={2025},
      eprint={2506.10185},
      archivePrefix={arXiv},
      primaryClass={quant-ph},
      url={https://arxiv.org/abs/2506.10185}, 
}

@article{sipser1996expander,
  title={Expander codes},
  author={Sipser, Michael and Spielman, Daniel A},
  journal={IEEE transactions on Information Theory},
  volume={42},
  number={6},
  pages={1710--1722},
  year={1996},
  publisher={IEEE},
  DOI={10.1109/18.556667},
  url={https://ieeexplore.ieee.org/document/556667}
}

@article{panteleev2021degenerate,
   title={Degenerate Quantum {LDPC} Codes With Good Finite Length Performance},
   volume={5},
   ISSN={2521-327X},
   url={http://dx.doi.org/10.22331/q-2021-11-22-585},
   DOI={10.22331/q-2021-11-22-585},
   journal={Quantum},
   publisher={Verein zur Forderung des Open Access Publizierens in den Quantenwissenschaften},
   author={Panteleev, Pavel and Kalachev, Gleb},
   year={2021},
   month=nov, pages={585} }

@article{panteleev2021quantum,
  title={Quantum {LDPC} codes with almost linear minimum distance},
  author={Panteleev, Pavel and Kalachev, Gleb},
  journal={IEEE Transactions on Information Theory},
  volume={68},
  number={1},
  pages={213--229},
  year={2021},
  publisher={IEEE},
  url={http://dx.doi.org/10.1109/TIT.2021.3119384},
  DOI={10.1109/tit.2021.3119384},
}

@inproceedings{panteleev2022asymptotically,
  title={Asymptotically good quantum and locally testable classical {LDPC} codes},
  author={Panteleev, Pavel and Kalachev, Gleb},
  booktitle={Proceedings of the 54th Annual ACM SIGACT Symposium on Theory of Computing},
  pages={375--388},
  year={2022},
  url = {https://doi.org/10.1145/3519935.3520017},
  doi = {10.1145/3519935.3520017}
}

@article{breuckmann2021balanced,
  title={Balanced product quantum codes},
  author={Breuckmann, Nikolas P and Eberhardt, Jens N},
  journal={IEEE Transactions on Information Theory},
  volume={67},
  number={10},
  pages={6653--6674},
  url={http://dx.doi.org/10.1109/TIT.2021.3097347},
  DOI={10.1109/tit.2021.3097347},
  year={2021},
  publisher={IEEE}
}

@inproceedings{leverrier2022quantum,
  title={Quantum {Tanner} codes},
  author={Leverrier, Anthony and Z{\'e}mor, Gilles},
  booktitle={2022 IEEE 63rd Annual Symposium on Foundations of Computer Science (FOCS)},
  pages={872--883},
  year={2022},
  organization={IEEE},
  doi = {10.1109/FOCS54457.2022.00117},
  url = {https://doi.ieeecomputersociety.org/10.1109/FOCS54457.2022.00117}
}

@inproceedings{dinur2023good,
author = {Dinur, Irit and Hsieh, Min-Hsiu and Lin, Ting-Chun and Vidick, Thomas},
title = {Good Quantum {LDPC} Codes with Linear Time Decoders},
year = {2023},
isbn = {9781450399135},
publisher = {Association for Computing Machinery},
address = {New York, NY, USA},
url = {https://doi.org/10.1145/3564246.3585101},
doi = {10.1145/3564246.3585101},
booktitle = {Proceedings of the 55th Annual ACM Symposium on Theory of Computing},
pages = {905–918},
numpages = {14},
location = {Orlando, FL, USA},
series = {STOC 2023}
}

@inproceedings{hastings2021fiber,
  title={Fiber bundle codes: breaking the {$\sqrt{n} \text{polylog}(n)$} barrier for quantum {LDPC} codes},
  author={Hastings, Matthew B and Haah, Jeongwan and O'Donnell, Ryan},
  booktitle={Proceedings of the 53rd Annual ACM SIGACT Symposium on Theory of Computing},
  pages={1276--1288},
  year={2021}
}

@article{naoyuki2020onsager,
  title = {Onsager's Scars in Disordered Spin Chains},
  author = {Shibata, Naoyuki and Yoshioka, Nobuyuki and Katsura, Hosho},
  journal = {Phys. Rev. Lett.},
  volume = {124},
  issue = {18},
  pages = {180604},
  numpages = {6},
  year = {2020},
  month = {May},
  publisher = {American Physical Society},
  doi = {10.1103/PhysRevLett.124.180604},
  url = {https://link.aps.org/doi/10.1103/PhysRevLett.124.180604}
}

@article{desaules2023weakschwinger,
   title={Weak ergodicity breaking in the {S}chwinger model},
   volume={107},
   ISSN={2469-9969},
   url={http://dx.doi.org/10.1103/PhysRevB.107.L201105},
   DOI={10.1103/physrevb.107.l201105},
   number={20},
   journal={Physical Review B},
   publisher={American Physical Society (APS)},
   author={Desaules, Jean-Yves and Banerjee, Debasish and Hudomal, Ana and Papić, Zlatko and Sen, Arnab and Halimeh, Jad C.},
   year={2023},
   month=may }

@article{zlatko2023briding,
   title={Bridging quantum criticality via many-body scarring},
   volume={107},
   ISSN={2469-9969},
   url={http://dx.doi.org/10.1103/PhysRevB.107.235108},
   DOI={10.1103/physrevb.107.235108},
   number={23},
   journal={Physical Review B},
   publisher={American Physical Society (APS)},
   author={Daniel, Aiden and Hallam, Andrew and Desaules, Jean-Yves and Hudomal, Ana and Su, Guo-Xian and Halimeh, Jad C. and Papić, Zlatko},
   year={2023},
   month=jun }

@misc{miao2025exactquantummanybodyscars,
      title={Exact Quantum Many-Body Scars in 2D Quantum Gauge Models}, 
      author={Yuan Miao and Linhao Li and Hosho Katsura and Masahito Yamazaki},
      year={2025},
      eprint={2505.21921},
      archivePrefix={arXiv},
      primaryClass={cond-mat.str-el},
      url={https://arxiv.org/abs/2505.21921}, 
}

@article{ge2024nonmesonic,
  title = {Nonmesonic Quantum Many-Body Scars in a 1D Lattice Gauge Theory},
  author = {Ge, Zi-Yong and Zhang, Yu-Ran and Nori, Franco},
  journal = {Phys. Rev. Lett.},
  volume = {132},
  issue = {23},
  pages = {230403},
  numpages = {8},
  year = {2024},
  month = {Jun},
  publisher = {American Physical Society},
  doi = {10.1103/PhysRevLett.132.230403},
  url = {https://link.aps.org/doi/10.1103/PhysRevLett.132.230403}
}

@misc{faugno2025density,
      title={Non-equilibirum physics of density-difference dependent {Hamiltonian}: Quantum Scarring from Emergent Chiral Symmetry}, 
      author={William N Faugno and Hosho Katsura and Tomoki Ozawa},
      year={2025},
      eprint={2503.05252},
      archivePrefix={arXiv},
      primaryClass={cond-mat.quant-gas},
      url={https://arxiv.org/abs/2503.05252}, 
}

@misc{sanada2024integrable,
      title={Towers of Quantum Many-body Scars from Integrable Boundary States}, 
      author={Kazuyuki Sanada and Yuan Miao and Hosho Katsura},
      year={2024},
      eprint={2411.01270},
      archivePrefix={arXiv},
      primaryClass={cond-mat.stat-mech},
      url={https://arxiv.org/abs/2411.01270}, 
}

@article{sanada2023multibody,
   title={Quantum many-body scars in spin models with multibody interactions},
   volume={108},
   ISSN={2469-9969},
   url={http://dx.doi.org/10.1103/PhysRevB.108.155102},
   DOI={10.1103/physrevb.108.155102},
   number={15},
   journal={Physical Review B},
   publisher={American Physical Society (APS)},
   author={Sanada, Kazuyuki and Miao, Yuan and Katsura, Hosho},
   year={2023},
   month=oct }

@article{tamura2022hopping,
   title={Quantum many-body scars of spinless fermions with density-assisted hopping in higher dimensions},
   volume={106},
   ISSN={2469-9969},
   url={http://dx.doi.org/10.1103/PhysRevB.106.144306},
   DOI={10.1103/physrevb.106.144306},
   number={14},
   journal={Physical Review B},
   publisher={American Physical Society (APS)},
   author={Tamura, Kensuke and Katsura, Hosho},
   year={2022},
   month=oct }

@article{udupa2023weak,
   title={Weak universality, quantum many-body scars, and anomalous infinite-temperature autocorrelations in a one-dimensional spin model with duality},
   volume={108},
   ISSN={2469-9969},
   url={http://dx.doi.org/10.1103/PhysRevB.108.214430},
   DOI={10.1103/physrevb.108.214430},
   number={21},
   journal={Physical Review B},
   publisher={American Physical Society (APS)},
   author={Udupa, Adithi and Sur, Samudra and Nandy, Sourav and Sen, Arnab and Sen, Diptiman},
   year={2023},
   month=dec }

@article{mestyan2025crosscap,
  title={Crosscap states with tunable entanglement as exact eigenstates of local spin chain {Hamiltonians}},
  author={M'arton Mesty'an and Bal{\'a}zs Pozsgay},
  journal={Journal of Physics A: Mathematical and Theoretical},
  year={2025},
  volume={58},
  url={https://api.semanticscholar.org/CorpusID:277150742}
}

@misc{mukherjee2025tensor,
      title={Symmetric tensor scars with tunable entanglement from volume to area law}, 
      author={Bhaskar Mukherjee and Christopher J. Turner and Marcin Szyniszewski and Arijeet Pal},
      year={2025},
      eprint={2501.14024},
      archivePrefix={arXiv},
      primaryClass={cond-mat.str-el},
      url={https://arxiv.org/abs/2501.14024}, 
}

@article{dong2023disorder,
   title={Disorder-tunable entanglement at infinite temperature},
   volume={9},
   ISSN={2375-2548},
   url={http://dx.doi.org/10.1126/sciadv.adj3822},
   DOI={10.1126/sciadv.adj3822},
   number={51},
   journal={Science Advances},
   publisher={American Association for the Advancement of Science (AAAS)},
   author={Dong, Hang and Desaules, Jean-Yves and Gao, Yu and Wang, Ning and Guo, Zexian and Chen, Jiachen and Zou, Yiren and Jin, Feitong and Zhu, Xuhao and Zhang, Pengfei and Li, Hekang and Wang, Zhen and Guo, Qiujiang and Zhang, Junxiang and Ying, Lei and Papić, Zlatko},
   year={2023},
   month=dec }

@article{srivatsa2023mobility,
   title={Mobility edges through inverted quantum many-body scarring},
   volume={108},
   ISSN={2469-9969},
   url={http://dx.doi.org/10.1103/PhysRevB.108.L100202},
   DOI={10.1103/physrevb.108.l100202},
   number={10},
   journal={Physical Review B},
   publisher={American Physical Society (APS)},
   author={Srivatsa, N. S. and Yarloo, Hadi and Moessner, Roderich and Nielsen, Anne E. B.},
   year={2023},
   month=sep }

@article{agarwal2023bell,
   title={Long-Range {Bell} States from Local Measurements and Many-Body Teleportation without Time Reversal},
   volume={130},
   ISSN={1079-7114},
   url={http://dx.doi.org/10.1103/PhysRevLett.130.020801},
   DOI={10.1103/physrevlett.130.020801},
   number={2},
   journal={Physical Review Letters},
   publisher={American Physical Society (APS)},
   author={Agarwal, Lakshya and Langlett, Christopher M. and Xu, Shenglong},
   year={2023},
   month=jan }

@Article{Deutsch1991quantum,
  title		= {Quantum statistical mechanics in a closed system},
  author	= {Deutsch, J. M.},
  journal	= {Phys. Rev. A},
  volume	= {43},
  issue		= {4},
  pages		= {2046--2049},
  numpages	= {0},
  year		= {1991},
  month		= {Feb},
  publisher	= {American Physical Society},
  doi		= {10.1103/PhysRevA.43.2046},
  url		= {https://link.aps.org/doi/10.1103/PhysRevA.43.2046}
}

@Article{Srednicki1994chaos,
  title		= {Chaos and quantum thermalization},
  author	= {Srednicki, Mark},
  journal	= {Phys. Rev. E},
  volume	= {50},
  issue		= {2},
  pages		= {888--901},
  numpages	= {0},
  year		= {1994},
  month		= {Aug},
  publisher	= {American Physical Society},
  doi		= {10.1103/PhysRevE.50.888},
  url		= {https://link.aps.org/doi/10.1103/PhysRevE.50.888}
}

@Article{	  Rigol2008thermalization,
  author	= {{Rigol}, Marcos and {Dunjko}, Vanja and {Olshanii},
		  Maxim},
  title		= "{Thermalization and its mechanism for generic isolated
		  quantum systems}",
  journal	= {Nature},
  year		= "2008",
  month		= "Apr",
  volume	= {452},
  number	= {7189},
  pages		= {854-858},
  doi		= {10.1038/nature06838}
}

@article{DAlessio2016quantum,
  author	= {{D'Alessio}, Luca and {Kafri}, Yariv and {Polkovnikov},
		  Anatoli and {Rigol}, Marcos},
  title		= "{From quantum chaos and eigenstate thermalization to
		  statistical mechanics and thermodynamics}",
  journal	= {Advances in Physics},
  year		= "2016",
  month		= "May",
  volume	= {65},
  number	= {3},
  pages		= {239-362},
  doi		= {10.1080/00018732.2016.1198134}
}

@misc{hashimoto2026construction,
      title={Construction of asymptotic quantum many-body scar states in the {SU}($N$) Hubbard model}, 
      author={Daiki Hashimoto and Masaya Kunimi and Tetsuro Nikuni},
      year={2026},
      eprint={2601.04640},
      archivePrefix={arXiv},
      primaryClass={cond-mat.stat-mech},
      url={https://arxiv.org/abs/2601.04640}, 
}

@article{batista2009spiral,
   title={Canted spiral: An exact ground state of {XXZ} zigzag spin ladders},
   volume={80},
   ISSN={1550-235X},
   url={http://dx.doi.org/10.1103/PhysRevB.80.180406},
   DOI={10.1103/physrevb.80.180406},
   number={18},
   journal={Physical Review B},
   publisher={American Physical Society (APS)},
   author={Batista, C. D.},
   year={2009},
   month=nov }

@article{desaules2024computer,
   title={Robust finite-temperature many-body scarring on a quantum computer},
   volume={110},
   ISSN={2469-9934},
   url={http://dx.doi.org/10.1103/PhysRevA.110.042606},
   DOI={10.1103/physreva.110.042606},
   number={4},
   journal={Physical Review A},
   publisher={American Physical Society (APS)},
   author={Desaules, Jean-Yves and Gustafson, Erik J. and Li, Andy C. Y. and Papić, Zlatko and Halimeh, Jad C.},
   year={2024},
   month=oct }

@article{chen2022pulse,
   title={Error-mitigated simulation of quantum many-body scars on quantum computers with pulse-level control},
   volume={4},
   ISSN={2643-1564},
   url={http://dx.doi.org/10.1103/PhysRevResearch.4.043027},
   DOI={10.1103/physrevresearch.4.043027},
   number={4},
   journal={Physical Review Research},
   publisher={American Physical Society (APS)},
   author={Chen, I-Chi and Burdick, Benjamin and Yao, Yongxin and Orth, Peter P. and Iadecola, Thomas},
   year={2022},
   month=oct }

@article{zhang2022superconducting,
   title={Many-body {H}ilbert space scarring on a superconducting processor},
   volume={19},
   ISSN={1745-2481},
   url={http://dx.doi.org/10.1038/s41567-022-01784-9},
   DOI={10.1038/s41567-022-01784-9},
   number={1},
   journal={Nature Physics},
   publisher={Springer Science and Business Media LLC},
   author={Zhang, Pengfei and Dong, Hang and Gao, Yu and Zhao, Liangtian and Hao, Jie and Desaules, Jean-Yves and Guo, Qiujiang and Chen, Jiachen and Deng, Jinfeng and Liu, Bobo and Ren, Wenhui and Yao, Yunyan and Zhang, Xu and Xu, Shibo and Wang, Ke and Jin, Feitong and Zhu, Xuhao and Zhang, Bing and Li, Hekang and Song, Chao and Wang, Zhen and Liu, Fangli and Papić, Zlatko and Ying, Lei and Wang, H. and Lai, Ying-Cheng},
   year={2022},
   month=oct, pages={120–125} }

@article{austinharris2025spinor,
   title={Observation of Ergodicity Breaking and Quantum Many-Body Scars in Spinor Gases},
   volume={134},
   ISSN={1079-7114},
   url={http://dx.doi.org/10.1103/PhysRevLett.134.113401},
   DOI={10.1103/physrevlett.134.113401},
   number={11},
   journal={Physical Review Letters},
   publisher={American Physical Society (APS)},
   author={Austin-Harris, J. O. and Rana, I. and Begg, S. E. and Binegar, C. and Bilitewski, T. and Liu, Y.},
   year={2025},
   month=mar }

@article{jepsen2022phantom,
   title={Long-lived phantom helix states in {Heisenberg} quantum magnets},
   volume={18},
   ISSN={1745-2481},
   url={http://dx.doi.org/10.1038/s41567-022-01651-7},
   DOI={10.1038/s41567-022-01651-7},
   number={8},
   journal={Nature Physics},
   publisher={Springer Science and Business Media LLC},
   author={Jepsen, Paul Niklas and Lee, Yoo Kyung `Eunice' and Lin, Hanzhen and Dimitrova, Ivana and Margalit, Yair and Ho, Wen Wei and Ketterle, Wolfgang},
   year={2022},
   month=jul, pages={899–904} }

@article{bluvstein2021controlling,
  title={Controlling quantum many-body dynamics in driven {R}ydberg atom arrays},
  author={Bluvstein, Dolev and Omran, Ahmed and Levine, Harry and Keesling, Alexander and Semeghini, Giulia and Ebadi, Sepehr and Wang, Tout T and Michailidis, Alexios A and Maskara, Nishad and Ho, Wen Wei and others},
  journal={Science},
  volume={371},
  number={6536},
  pages={1355--1359},
  year={2021},
  publisher={American Association for the Advancement of Science},
  url={http://dx.doi.org/10.1126/science.abg2530},
  DOI={10.1126/science.abg2530},
}

@article{bluvstein2022processor,
   title={A quantum processor based on coherent transport of entangled atom arrays},
   volume={604},
   ISSN={1476-4687},
   url={http://dx.doi.org/10.1038/s41586-022-04592-6},
   DOI={10.1038/s41586-022-04592-6},
   number={7906},
   journal={Nature},
   publisher={Springer Science and Business Media LLC},
   author={Bluvstein, Dolev and Levine, Harry and Semeghini, Giulia and Wang, Tout T. and Ebadi, Sepehr and Kalinowski, Marcin and Keesling, Alexander and Maskara, Nishad and Pichler, Hannes and Greiner, Markus and Vuletić, Vladan and Lukin, Mikhail D.},
   year={2022},
   month=apr, pages={451–456} }

@article{kao2021pumping,
   title={Topological pumping of a 1D dipolar gas into strongly correlated prethermal states},
   volume={371},
   ISSN={1095-9203},
   url={http://dx.doi.org/10.1126/science.abb4928},
   DOI={10.1126/science.abb4928},
   number={6526},
   journal={Science},
   publisher={American Association for the Advancement of Science (AAAS)},
   author={Kao, Wil and Li, Kuan-Yu and Lin, Kuan-Yu and Gopalakrishnan, Sarang and Lev, Benjamin L.},
   year={2021},
   month=jan, pages={296–300} }

@article{su2023bose,
  title = {Observation of many-body scarring in a {Bose}-{Hubbard} quantum simulator},
  author = {Su, Guo-Xian and Sun, Hui and Hudomal, Ana and Desaules, Jean-Yves and Zhou, Zhao-Yu and Yang, Bing and Halimeh, Jad C. and Yuan, Zhen-Sheng and Papi\ifmmode \acute{c}\else \'{c}\fi{}, Zlatko and Pan, Jian-Wei},
  journal = {Phys. Rev. Res.},
  volume = {5},
  issue = {2},
  pages = {023010},
  numpages = {13},
  year = {2023},
  month = {Apr},
  publisher = {American Physical Society},
  doi = {10.1103/PhysRevResearch.5.023010},
  url = {https://link.aps.org/doi/10.1103/PhysRevResearch.5.023010}
}

@article{Serbyn-Papic2021_review,
	author = {Serbyn, Maksym and Abanin, Dmitry A. and Papi{\'c}, Zlatko},
	date = {2021/06/01},
	doi = {10.1038/s41567-021-01230-2},
	id = {Serbyn2021},
	isbn = {1745-2481},
	journal = {Nature Physics},
	number = {6},
	pages = {675--685},
	title = {Quantum many-body scars and weak breaking of ergodicity},
	url = {https://doi.org/10.1038/s41567-021-01230-2},
	volume = {17},
	year = {2021}}

@article{Moudgalya-Regnault2021_review,
	doi = {10.1088/1361-6633/ac73a0},
	url = {https://doi.org/10.1088/1361-6633/ac73a0},
	year = 2022,
	month = {jul},
	publisher = {{IOP} Publishing},
	volume = {85},
	number = {8},
	pages = {086501},
	author = {Sanjay Moudgalya and B Andrei Bernevig and Nicolas Regnault},
	title = {Quantum many-body scars and {H}ilbert space fragmentation: a review of exact results},
	journal = {Reports on Progress in Physics},
}

@article{Chandran-Moessner2022_review,
author = {Chandran, Anushya and Iadecola, Thomas and Khemani, Vedika and Moessner, Roderich},
title = {Quantum Many-Body Scars: A Quasiparticle Perspective},
journal = {Annual Review of Condensed Matter Physics},
volume = {14},
number = {1},
pages = {443-469},
year = {2023},
doi = {10.1146/annurev-conmatphys-031620-101617},
URL = {https://doi.org/10.1146/annurev-conmatphys-031620-101617},
}

@article{moudgalya2018nonintegrable,
  title = {Exact excited states of nonintegrable models},
  author = {Moudgalya, Sanjay and Rachel, Stephan and Bernevig, B. Andrei and Regnault, Nicolas},
  journal = {Phys. Rev. B},
  volume = {98},
  issue = {23},
  pages = {235155},
  numpages = {31},
  year = {2018},
  month = {Dec},
  publisher = {American Physical Society},
  doi = {10.1103/PhysRevB.98.235155},
  url = {https://link.aps.org/doi/10.1103/PhysRevB.98.235155}
}

@article{moudgalya2018entanglement,
  title = {Entanglement of exact excited states of {Affleck}-{Kennedy}-{Lieb}-{Tasaki} models: Exact results, many-body scars, and violation of the strong eigenstate thermalization hypothesis},
  author = {Moudgalya, Sanjay and Regnault, Nicolas and Bernevig, B. Andrei},
  journal = {Phys. Rev. B},
  volume = {98},
  issue = {23},
  pages = {235156},
  numpages = {43},
  year = {2018},
  month = {Dec},
  publisher = {American Physical Society},
  doi = {10.1103/PhysRevB.98.235156},
  url = {https://link.aps.org/doi/10.1103/PhysRevB.98.235156}
}

@article{ren2021quasisymmetry,
  title = {Quasisymmetry Groups and Many-Body Scar Dynamics},
  author = {Ren, Jie and Liang, Chenguang and Fang, Chen},
  journal = {Phys. Rev. Lett.},
  volume = {126},
  issue = {12},
  pages = {120604},
  numpages = {6},
  year = {2021},
  month = {Mar},
  publisher = {American Physical Society},
  doi = {10.1103/PhysRevLett.126.120604},
  url = {https://link.aps.org/doi/10.1103/PhysRevLett.126.120604}
}

@article{tang2022multimagnon,
   title={Multimagnon quantum many-body scars from tensor operators},
   volume={4},
   ISSN={2643-1564},
   url={http://dx.doi.org/10.1103/PhysRevResearch.4.043006},
   DOI={10.1103/physrevresearch.4.043006},
   number={4},
   journal={Physical Review Research},
   publisher={American Physical Society (APS)},
   author={Tang, Long-Hin and O’Dea, Nicholas and Chandran, Anushya},
   year={2022},
   month=oct }

@misc{gioia2025distinct,
      title={Distinct Types of Parent {Hamiltonians} for Quantum States: Insights from the ${W}$ State as a Quantum Many-Body Scar}, 
      author={Lei Gioia and Sanjay Moudgalya and Olexei I. Motrunich},
      year={2025},
      eprint={2510.24713},
      archivePrefix={arXiv},
      primaryClass={quant-ph},
      url={https://arxiv.org/abs/2510.24713}, 
}

@article{keita2023projector2,
  title = {Fractionalization paves the way to local projector embeddings of quantum many-body scars},
  author = {Omiya, Keita and Muller, Markus},
  journal = {Phys. Rev. B},
  volume = {108},
  issue = {5},
  pages = {054412},
  numpages = {18},
  year = {2023},
  month = {Aug},
  publisher = {American Physical Society},
  doi = {10.1103/PhysRevB.108.054412},
  url = {https://link.aps.org/doi/10.1103/PhysRevB.108.054412}
}

@misc{Mark2025observation,
      title={Observation of ballistic plasma and memory in high-energy gauge theory dynamics},
      author={Daniel K. Mark and Federica M. Surace and Thomas Schuster and Adam L. Shaw and Wenjie Gong and Soonwon Choi and Manuel Endres},
      year={2025},
      eprint={2510.11679},
      archivePrefix={arXiv},
      primaryClass={quant-ph},
      url={https://arxiv.org/abs/2510.11679}, 
}

@article{Mark2020Eta,
  title = {$\ensuremath{\eta}$-pairing states as true scars in an extended {Hubbard} model},
  author = {Mark, Daniel K. and Motrunich, Olexei I.},
  journal = {Phys. Rev. B},
  volume = {102},
  issue = {7},
  pages = {075132},
  numpages = {18},
  year = {2020},
  month = {Aug},
  publisher = {American Physical Society},
  doi = {10.1103/PhysRevB.102.075132},
  url = {https://link.aps.org/doi/10.1103/PhysRevB.102.075132}
}

@article{Kerschbaumer2025quantum,
  title = {Quantum Many-Body Scars beyond the {PXP} Model in {R}ydberg Simulators},
  author = {Kerschbaumer, Aron and Ljubotina, Marko and Serbyn, Maksym and Desaules, Jean-Yves},
  journal = {Phys. Rev. Lett.},
  volume = {134},
  issue = {16},
  pages = {160401},
  numpages = {9},
  year = {2025},
  month = {Apr},
  publisher = {American Physical Society},
  doi = {10.1103/PhysRevLett.134.160401},
  url = {https://link.aps.org/doi/10.1103/PhysRevLett.134.160401}
}

@misc{Kerschbaumer2025discrete,
      title={Discrete solitons in {R}ydberg atom chains}, 
      author={Aron Kerschbaumer and Jean-Yves Desaules and Marko Ljubotina and Maksym Serbyn},
      year={2025},
      eprint={2507.13196},
      archivePrefix={arXiv},
      primaryClass={quant-ph},
      url={https://arxiv.org/abs/2507.13196}, 
}

@article{Schecter2019weak,
  title = {Weak Ergodicity Breaking and Quantum Many-Body Scars in Spin-1 ${XY}$ Magnets},
  author = {Schecter, Michael and Iadecola, Thomas},
  journal = {Phys. Rev. Lett.},
  volume = {123},
  issue = {14},
  pages = {147201},
  numpages = {5},
  year = {2019},
  month = {Oct},
  publisher = {American Physical Society},
  doi = {10.1103/PhysRevLett.123.147201},
  url = {https://link.aps.org/doi/10.1103/PhysRevLett.123.147201}
}

@article{pakrouski2020many,
	title        = {Many-Body Scars as a Group Invariant Sector of {Hilbert} Space},
	author       = {Pakrouski, K. and Pallegar, P. N. and Popov, F. K. and Klebanov, I. R.},
	year         = 2020,
	month        = {Dec},
	journal      = {Phys. Rev. Lett.},
	publisher    = {American Physical Society},
	volume       = 125,
	pages        = 230602,
	doi          = {10.1103/PhysRevLett.125.230602},
	url          = {https://link.aps.org/doi/10.1103/PhysRevLett.125.230602},
	issue        = 23,
	numpages     = 6
}

@article{Ivanov2024volume,
  title = {Volume-Entangled Exact Scar States in the {P}{X}{P} and Related Models in Any Dimension},
  author = {Ivanov, Andrew N. and Motrunich, Olexei I.},
  journal = {Phys. Rev. Lett.},
  volume = {134},
  issue = {5},
  pages = {050403},
  numpages = {6},
  year = {2025},
  month = {Feb},
  publisher = {American Physical Society},
  doi = {10.1103/PhysRevLett.134.050403},
  url = {https://link.aps.org/doi/10.1103/PhysRevLett.134.050403}
}

@article{mohapatra2024exact,
  title   = {Exact Volume-Law Entangled Zero-Energy Eigenstates in a Large Class of Spin Models},
  author  = {Mohapatra, Sashikanta and Moudgalya, Sanjay and Balram, Ajit C.},
  journal = {Physical Review Letters},
  volume  = {134},
  number  = {21},
  pages   = {210403},
  year    = {2025},
  month   = may,
  doi     = {10.1103/PhysRevLett.134.210403},
  url     = {https://doi.org/10.1103/PhysRevLett.134.210403}
}

@article{mark2020unified,
	title        = {Unified structure for exact towers of scar states in the {Affleck}-{Kennedy}-{Lieb}-{Tasaki} and other models},
	author       = {Mark, Daniel K. and Lin, Cheng-Ju and Motrunich, Olexei I.},
	year         = 2020,
	month        = {May},
	journal      = {Phys. Rev. B},
	publisher    = {American Physical Society},
	volume       = 101,
	pages        = 195131,
	doi          = {10.1103/PhysRevB.101.195131},
	url          = {https://link.aps.org/doi/10.1103/PhysRevB.101.195131},
	issue        = 19,
	numpages     = 17
}

@article{pakrouski2021group,
	title        = {Group theoretic approach to many-body scar states in fermionic lattice models},
	author       = {Pakrouski, K. and Pallegar, P. N. and Popov, F. K. and Klebanov, I. R.},
	year         = 2021,
	month        = {Dec},
	journal      = {Phys. Rev. Research},
	publisher    = {American Physical Society},
	volume       = 3,
	pages        = {043156},
	doi          = {10.1103/PhysRevResearch.3.043156},
	url          = {https://link.aps.org/doi/10.1103/PhysRevResearch.3.043156},
	issue        = 4,
	numpages     = 20
}

@article{ren2022deformed,
	title        = {Deformed symmetry structures and quantum many-body scar subspaces},
	author       = {Ren, Jie and Liang, Chenguang and Fang, Chen},
	year         = 2022,
	month        = {Feb},
	journal      = {Phys. Rev. Research},
	publisher    = {American Physical Society},
	volume       = 4,
	pages        = {013155},
	doi          = {10.1103/PhysRevResearch.4.013155},
	url          = {https://link.aps.org/doi/10.1103/PhysRevResearch.4.013155},
	issue        = 1,
	numpages     = 26
}

@article{Moudgalya2020Large,
  title = {Large classes of quantum scarred {Hamiltonians} from matrix product states},
  author = {Moudgalya, Sanjay and O'Brien, Edward and Bernevig, B. Andrei and Fendley, Paul and Regnault, Nicolas},
  journal = {Phys. Rev. B},
  volume = {102},
  issue = {8},
  pages = {085120},
  numpages = {19},
  year = {2020},
  month = {Aug},
  publisher = {American Physical Society},
  doi = {10.1103/PhysRevB.102.085120},
  url = {https://link.aps.org/doi/10.1103/PhysRevB.102.085120}
}

@article{moudgalya2020eta,
	title        = {{$\ensuremath{\eta}$-pairing in {Hubbard} models: From spectrum generating algebras to quantum many-body scars}},
	author       = {Moudgalya, Sanjay and Regnault, Nicolas and Bernevig, B. Andrei},
	year         = 2020,
	month        = {Aug},
	journal      = {Phys. Rev. B},
	publisher    = {American Physical Society},
	volume       = 102,
	pages        = {085140},
	doi          = {10.1103/PhysRevB.102.085140},
	url          = {https://link.aps.org/doi/10.1103/PhysRevB.102.085140},
	issue        = 8,
	numpages     = 12
}

@article{moudgalya2023exhaustive,
  title = {Exhaustive Characterization of Quantum Many-Body Scars Using Commutant Algebras},
  author = {Moudgalya, Sanjay and Motrunich, Olexei I.},
  journal = {Phys. Rev. X},
  volume = {14},
  issue = {4},
  pages = {041069},
  numpages = {49},
  year = {2024},
  month = {Dec},
  publisher = {American Physical Society},
  doi = {10.1103/PhysRevX.14.041069},
  url = {https://link.aps.org/doi/10.1103/PhysRevX.14.041069}
}

@article{odea2020from,
  title = {From tunnels to towers: Quantum scars from {L}ie algebras and $q$-deformed {L}ie algebras},
  author = {O'Dea, Nicholas and Burnell, Fiona and Chandran, Anushya and Khemani, Vedika},
  journal = {Phys. Rev. Research},
  volume = {2},
  issue = {4},
  pages = {043305},
  numpages = {30},
  year = {2020},
  month = {Dec},
  publisher = {American Physical Society},
  doi = {10.1103/PhysRevResearch.2.043305},
  url = {https://link.aps.org/doi/10.1103/PhysRevResearch.2.043305}
}

@article{Shiraishi_2017,
  title = {Systematic Construction of Counterexamples to the Eigenstate Thermalization Hypothesis},
  author = {Shiraishi, Naoto and Mori, Takashi},
  journal = {Phys. Rev. Lett.},
  volume = {119},
  issue = {3},
  pages = {030601},
  numpages = {6},
  year = {2017},
  month = {Jul},
  publisher = {American Physical Society},
  doi = {10.1103/PhysRevLett.119.030601},
  url = {https://link.aps.org/doi/10.1103/PhysRevLett.119.030601}
}

@article{Turner_2018weak,
  author = {{Turner}, C.~J. and {Michailidis}, A.~A. and
            {Abanin}, D.~A. and {Serbyn}, M. and {Papi{\'c}}, Z.},
  title = "{Weak ergodicity breaking from quantum many-body scars}",
  journal = {Nature Physics},
  year = 2018,
  month = may,
  volume = 14,
  number = 7,
  pages = {745-749},
  doi = {10.1038/s41567-018-0137-5}
}

@article{Turner_2018quantum,
  title = {Quantum scarred eigenstates in a {R}ydberg atom chain: Entanglement, breakdown of thermalization, and stability to perturbations},
  author = {Turner, C. J. and Michailidis, A. A. and Abanin, D. A. and Serbyn, M. and Papi\ifmmode \acute{c}\else \'{c}\fi{}, Z.},
  journal = {Phys. Rev. B},
  volume = {98},
  issue = {15},
  pages = {155134},
  numpages = {23},
  year = {2018},
  month = {Oct},
  publisher = {American Physical Society},
  doi = {10.1103/PhysRevB.98.155134},
  url = {https://link.aps.org/doi/10.1103/PhysRevB.98.155134}
}

@article{Khemani_2019,
  title = {Signatures of integrability in the dynamics of {R}ydberg-blockaded chains},
  author = {Khemani, Vedika and Laumann, Chris R. and Chandran, Anushya},
  journal = {Phys. Rev. B},
  volume = {99},
  issue = {16},
  pages = {161101},
  numpages = {6},
  year = {2019},
  month = {Apr},
  publisher = {American Physical Society},
  doi = {10.1103/PhysRevB.99.161101},
  url = {https://link.aps.org/doi/10.1103/PhysRevB.99.161101}
}

@article{Lin_2019,
  title = {Exact Quantum Many-Body Scar States in the {R}ydberg-Blockaded Atom Chain},
  author = {Lin, Cheng-Ju and Motrunich, Olexei I.},
  journal = {Phys. Rev. Lett.},
  volume = {122},
  issue = {17},
  pages = {173401},
  numpages = {5},
  year = {2019},
  month = {Apr},
  publisher = {American Physical Society},
  doi = {10.1103/PhysRevLett.122.173401},
  url = {https://link.aps.org/doi/10.1103/PhysRevLett.122.173401}
}

@misc{Ivanov2025exact,
      title={Many exact area-law scar eigenstates in the nonintegrable {PXP} and related models}, 
      author={Andrew N. Ivanov and Olexei I. Motrunich},
      eprint={2503.16327},
      archivePrefix={arXiv},
      year = {2025}
}

@article{Choi_2019,
  title = {Emergent {SU}(2) Dynamics and Perfect Quantum Many-Body Scars},
  author = {Choi, Soonwon and Turner, Christopher J. and Pichler, Hannes and Ho, Wen Wei and Michailidis, Alexios A. and Papi\ifmmode \acute{c}\else \'{c}\fi{}, Zlatko and Serbyn, Maksym and Lukin, Mikhail D. and Abanin, Dmitry A.},
  journal = {Phys. Rev. Lett.},
  volume = {122},
  issue = {22},
  pages = {220603},
  numpages = {6},
  year = {2019},
  month = {Jun},
  publisher = {American Physical Society},
  doi = {10.1103/PhysRevLett.122.220603},
  url = {https://link.aps.org/doi/10.1103/PhysRevLett.122.220603}
}

@article{Surace_2020,
  title = {Lattice Gauge Theories and String Dynamics in {R}ydberg Atom Quantum Simulators},
  author = {Surace, Federica M. and Mazza, Paolo P. and Giudici, Giuliano and Lerose, Alessio and Gambassi, Andrea and Dalmonte, Marcello},
  journal = {Phys. Rev. X},
  volume = {10},
  issue = {2},
  pages = {021041},
  numpages = {14},
  year = {2020},
  month = {May},
  publisher = {American Physical Society},
  doi = {10.1103/PhysRevX.10.021041},
  url = {https://link.aps.org/doi/10.1103/PhysRevX.10.021041}
}

@article{Iadecola_2019,
  title = {Quantum many-body scars from magnon condensation},
  author = {Iadecola, Thomas and Schecter, Michael and Xu, Shenglong},
  journal = {Phys. Rev. B},
  volume = {100},
  issue = {18},
  pages = {184312},
  numpages = {12},
  year = {2019},
  month = {Nov},
  publisher = {American Physical Society},
  doi = {10.1103/PhysRevB.100.184312},
  url = {https://link.aps.org/doi/10.1103/PhysRevB.100.184312}
}

@article{giudici2023unraveling,
   title={Unraveling {PXP} Many-Body Scars through {F}loquet Dynamics},
   volume={133},
   ISSN={1079-7114},
   url={http://dx.doi.org/10.1103/PhysRevLett.133.190404},
   DOI={10.1103/physrevlett.133.190404},
   number={19},
   journal={Physical Review Letters},
   publisher={American Physical Society (APS)},
   author={Giudici, Giuliano and Surace, Federica Maria and Pichler, Hannes},
   year={2024},
   month=nov }

@incollection{Papic_2022,
  author       = {Zlatko Papi{\'c}},
  title        = {Weak Ergodicity Breaking Through the Lens of Quantum Entanglement},
  booktitle    = {Entanglement in Spin Chains},
  editor       = {Abolfazl Bayat and Sougato Bose and Henrik Johannesson},
  publisher    = {Springer International Publishing},
  address      = {Cham},
  year         = {2022},
  pages        = {341--395},
  doi          = {10.1007/978-3-031-03998-0_13},
  isbn         = {978-3-031-03998-0},
  url          = {https://link.springer.com/chapter/10.1007/978-3-031-03998-0_13},
  eprint       = {2108.03460},
  archivePrefix= {arXiv},
  primaryClass = {cond-mat.stat-mech}
}

@ARTICLE{kunimi2025systematic,
       author = {{Kunimi}, Masaya and {Kato}, Yusuke and {Katsura}, Hosho},
        title = "{Systematic construction of asymptotic quantum many-body scar states and their relation to supersymmetric quantum mechanics}",
      journal = {arXiv e-prints},
         year = 2025,
        month = may,
          eid = {arXiv:2505.04853},
        pages = {arXiv:2505.04853},
          doi = {10.48550/arXiv.2505.04853},
archivePrefix = {arXiv},
       eprint = {2505.04853},
 primaryClass = {cond-mat.stat-mech},
       adsurl = {https://ui.adsabs.harvard.edu/abs/2025arXiv250504853K}
}

@article{Langlett_2022,
  title = {Rainbow scars: From area to volume law},
  author = {Langlett, Christopher M. and Yang, Zhi-Cheng and Wildeboer, Julia and Gorshkov, Alexey V. and Iadecola, Thomas and Xu, Shenglong},
  journal = {Phys. Rev. B},
  volume = {105},
  issue = {6},
  pages = {L060301},
  numpages = {6},
  year = {2022},
  month = {Feb},
  publisher = {American Physical Society},
  doi = {10.1103/PhysRevB.105.L060301},
  url = {https://link.aps.org/doi/10.1103/PhysRevB.105.L060301}
}

@article{odea2025entanglement,
  title   = {Entanglement Oscillations from Many-Body Quantum Scars},
  author  = {O'Dea, Nicholas and Sriram, Adithya},
  journal = {Physical Review Letters},
  volume  = {134},
  number  = {21},
  pages   = {210402},
  year    = {2025},
  month   = may,
  doi     = {10.1103/PhysRevLett.134.210402},
  url     = {https://doi.org/10.1103/PhysRevLett.134.210402}
}

@article{qiranard2017,
	title        = {Determining a local {H}amiltonian from a single eigenstate},
	author       = {Qi, Xiao-Liang and Ranard, Daniel},
	year         = 2019,
	month        = jul,
	journal      = {{Quantum}},
	publisher    = {{Verein zur F{\"{o}}rderung des Open Access Publizierens in den Quantenwissenschaften}},
	volume       = 3,
	pages        = 159,
	doi          = {10.22331/q-2019-07-08-159},
	issn         = {2521-327X},
	url          = {https://doi.org/10.22331/q-2019-07-08-159}
}

@article{chertkovclark2018,
	title        = {Engineering topological models with a general-purpose symmetry-to-{Hamiltonian} approach},
	author       = {Chertkov, Eli and Villalonga, Benjamin and Clark, Bryan K.},
	year         = 2020,
	month        = {Jun},
	journal      = {Phys. Rev. Research},
	publisher    = {American Physical Society},
	volume       = 2,
	pages        = {023348},
	doi          = {10.1103/PhysRevResearch.2.023348},
	url          = {https://link.aps.org/doi/10.1103/PhysRevResearch.2.023348},
	issue        = 2,
	numpages     = 24
}

@ARTICLE{gotta2025open,
       author = {{Gotta}, Lorenzo},
        title = "{Open-system quantum many-body scars: a theory}",
      journal = {arXiv e-prints},
         year = 2025,
        month = sep,
          eid = {arXiv:2509.18023},
        pages = {arXiv:2509.18023},
          doi = {10.48550/arXiv.2509.18023},
archivePrefix = {arXiv},
       eprint = {2509.18023},
 primaryClass = {quant-ph},
       adsurl = {https://ui.adsabs.harvard.edu/abs/2025arXiv250918023G}
}

@misc{gioia2024wstateuniqueground,
      title={{$W$} state is not the unique ground state of any local {Hamiltonian}}, 
      author={Lei Gioia and Ryan Thorngren},
      year={2024},
      eprint={2310.10716},
      archivePrefix={arXiv},
      primaryClass={cond-mat.str-el},
      url={https://arxiv.org/abs/2310.10716}, 
}

@article{bernien2017probing,
  title     = {Probing many-body dynamics on a 51-atom quantum simulator},
  author    = {Bernien, Hannes and Schwartz, Sylvain and Keesling, Alexander and Levine, Harry and Omran, Ahmed and Pichler, Hannes and Choi, Soonwon and Zibrov, Alexander S and Endres, Manuel and Greiner, Markus and Vuleti{\'c}, Vladan and Lukin, Mikhail D},
  journal   = {Nature},
  volume    = {551},
  number    = {7682},
  pages     = {579--584},
  year      = {2017},
  publisher = {Nature Publishing Group},
  doi       = {10.1038/nature24622},
  url       = {https://doi.org/10.1038/nature24622}
}

@article{haffner2005scalable,
  title     = {Scalable multiparticle entanglement of trapped ions},
  author    = {H{\"a}ffner, H and H{\"a}nsel, W and Roos, C F and Benhelm, J and Chek-al-kar, D and Chwalla, M and K{\"o}rber, T and Rapol, U D and Riebe, M and Schmidt, P O and Becher, C and G{\"u}hne, O and D{\"u}r, W and Blatt, R},
  journal   = {Nature},
  volume    = {438},
  number    = {7068},
  pages     = {643--646},
  year      = {2005},
  publisher = {Nature Publishing Group},
  doi       = {10.1038/nature04279},
  url       = {https://doi.org/10.1038/nature04279}
}

@article{song2017ten,
  title     = {10-qubit entanglement and parallel logic operations with a superconducting circuit},
  author    = {Song, Chao and Xu, Ke and Liu, Wuxin and Yang, Chui-Ping and Zheng, Shi-Biao and Deng, Hui and Xie, Qing and Huang, Kui and Guo, Qi and Zhang, Luyan and Zhang, Peng and Xu, Dongning and Zheng, Ding and Deng, Chenyong and Rong, Heng and You, JQ and Wang, Zhen and Chen, Yu-Ao and Pan, Jian-Wei},
  journal   = {Physical Review Letters},
  volume    = {119},
  number    = {18},
  pages     = {180511},
  year      = {2017},
  publisher = {APS},
  doi       = {10.1103/PhysRevLett.119.180511},
  url       = {https://doi.org/10.1103/PhysRevLett.119.180511}
}

@article{eibl2004experimental,
  title     = {Experimental realization of a three-qubit entangled {W} state},
  author    = {Eibl, M and Kiesel, N and Bourennane, M and Kurtsiefer, C and Weinfurter, H},
  journal   = {Physical Review Letters},
  volume    = {92},
  number    = {7},
  pages     = {077901},
  year      = {2004},
  publisher = {APS},
  doi       = {10.1103/PhysRevLett.92.077901},
  url       = {https://doi.org/10.1103/PhysRevLett.92.077901}
}

@article{RevModPhys.93.045003,
  title = {Matrix product states and projected entangled pair states: Concepts, symmetries, theorems},
  author = {Cirac, J. Ignacio and P\'erez-Garc\'{\i}a, David and Schuch, Norbert and Verstraete, Frank},
  journal = {Rev. Mod. Phys.},
  volume = {93},
  issue = {4},
  pages = {045003},
  numpages = {65},
  year = {2021},
  month = {Dec},
  publisher = {American Physical Society},
  doi = {10.1103/RevModPhys.93.045003},
  url = {https://link.aps.org/doi/10.1103/RevModPhys.93.045003}
}

@article{PhysRevA.98.062335,
  title = {Entanglement robustness against particle loss in multiqubit systems},
  author = {Neven, A. and Martin, J. and Bastin, T.},
  journal = {Phys. Rev. A},
  volume = {98},
  issue = {6},
  pages = {062335},
  numpages = {6},
  year = {2018},
  month = {Dec},
  publisher = {American Physical Society},
  doi = {10.1103/PhysRevA.98.062335},
  url = {https://link.aps.org/doi/10.1103/PhysRevA.98.062335}
}

@article{PhysRevA.62.062314,
  title = {Three qubits can be entangled in two inequivalent ways},
  author = {D\"ur, W. and Vidal, G. and Cirac, J. I.},
  journal = {Phys. Rev. A},
  volume = {62},
  issue = {6},
  pages = {062314},
  numpages = {12},
  year = {2000},
  month = {Nov},
  publisher = {American Physical Society},
  doi = {10.1103/PhysRevA.62.062314},
  url = {https://link.aps.org/doi/10.1103/PhysRevA.62.062314}
}

@article{PhysRevA.75.052109,
  title = {Evolution of a quantum spin system to its ground state: Role of entanglement and interaction symmetry},
  author = {Yuan, Shengjun and Katsnelson, Mikhail I. and De Raedt, Hans},
  journal = {Phys. Rev. A},
  volume = {75},
  issue = {5},
  pages = {052109},
  numpages = {10},
  year = {2007},
  month = {May},
  publisher = {American Physical Society},
  doi = {10.1103/PhysRevA.75.052109},
  url = {https://link.aps.org/doi/10.1103/PhysRevA.75.052109}
}

@article{kieferova2024logdepth,
  title     = {State preparation by shallow circuits using feed forward},
  author    = {Kieferov\'a, Benedikt and Wang, David S. and Janzing, Dominik},
  journal   = {Quantum},
  volume    = {8},
  pages     = {1552},
  year      = {2024},
  doi       = {10.22331/q-2024-12-09-1552},
  url       = {https://quantum-journal.org/papers/q-2024-12-09-1552/}
}

@article{Gotta2023AsymptoticScars,
  title         = {Asymptotic Quantum Many-Body Scars},
  author        = {Gotta, Lorenzo and Moudgalya, Sanjay and Mazza, Leonardo},
  journal       = {Phys. Rev. Lett.},
  volume        = {131},
  number        = {19},
  pages         = {190401},
  year          = {2023},
  publisher     = {American Physical Society},
  doi           = {10.1103/PhysRevLett.131.190401},
  url           = {https://doi.org/10.1103/PhysRevLett.131.190401},
  eprint        = {2303.05407},
  archivePrefix = {arXiv}
}

@article{Kunimi2023AsymptoticScarsDMI,
  title = {Proposal for simulating quantum spin models with the {D}zyaloshinskii-{M}oriya interaction using {R}ydberg atoms and the construction of asymptotic quantum many-body scar states},
  author = {Kunimi, Masaya and Tomita, Takafumi and Katsura, Hosho and Kato, Yusuke},
  journal = {Phys. Rev. A},
  volume = {110},
  issue = {4},
  pages = {043312},
  numpages = {18},
  year = {2024},
  month = {Oct},
  publisher = {American Physical Society},
  doi = {10.1103/PhysRevA.110.043312},
  url = {https://link.aps.org/doi/10.1103/PhysRevA.110.043312}
}

@article{Desaules2022SchwingerScars,
   title={Prominent quantum many-body scars in a truncated {S}chwinger model},
   volume={107},
   ISSN={2469-9969},
   url={http://dx.doi.org/10.1103/PhysRevB.107.205112},
   DOI={10.1103/physrevb.107.205112},
   number={20},
   journal={Physical Review B},
   publisher={American Physical Society (APS)},
   author={Desaules, Jean-Yves and Hudomal, Ana and Banerjee, Debasish and Sen, Arnab and Papić, Zlatko and Halimeh, Jad C.},
   year={2023},
   month=may }

@article{chiba2024exact,
  title = {Exact Thermal Eigenstates of Nonintegrable Spin Chains at Infinite Temperature},
  author = {Chiba, Yuuya and Yoneta, Yasushi},
  journal = {Phys. Rev. Lett.},
  volume = {133},
  issue = {17},
  pages = {170404},
  numpages = {6},
  year = {2024},
  month = {Oct},
  publisher = {American Physical Society},
  doi = {10.1103/PhysRevLett.133.170404},
  url = {https://link.aps.org/doi/10.1103/PhysRevLett.133.170404}
}

@article{Perez-Garcia2007,
  author       = {P{\'e}rez-Garc{\'\i}a, David and Verstraete, Frank and Wolf, Michael M. and Cirac, J. Ignacio},
  title        = {Matrix Product State Representations},
  journal      = {Quantum Information \& Computation},
  volume       = {7},
  number       = {5-6},
  pages        = {401--430},
  year         = {2007},
  doi          = {10.26421/QIC7.5-6-1},
  eprint       = {quant-ph/0608197},
  archivePrefix= {arXiv},
  primaryClass = {quant-ph}
}

@article{AffleckKennedyLiebTasaki1987,
  author  = {Affleck, Ian and Kennedy, Tom and Lieb, Elliott H. and Tasaki, Hal},
  title   = {Rigorous results on valence-bond ground states in antiferromagnets},
  journal = {Physical Review Letters},
  volume  = {59},
  number  = {7},
  pages   = {799--802},
  year    = {1987},
  month   = aug,
  doi     = {10.1103/PhysRevLett.59.799}
}

@article{FannesNachtergaeleWerner1992,
  author  = {Fannes, M. and Nachtergaele, B. and Werner, R. F.},
  title   = {Finitely Correlated States on Quantum Spin Chains},
  journal = {Communications in Mathematical Physics},
  volume  = {144},
  number  = {3},
  pages   = {443--490},
  year    = {1992},
  doi     = {10.1007/BF02099178}
}

@article{Dicke1954Coherence,
  author    = {R. H. Dicke},
  title     = {Coherence in Spontaneous Radiation Processes},
  journal   = {Physical Review},
  year      = {1954},
  volume    = {93},
  pages     = {99--110},
  doi       = {10.1103/PhysRev.93.99},
  url       = {https://link.aps.org/doi/10.1103/PhysRev.93.99}
}

\end{document}